\documentclass[10pt,twocolumn,twoside]{IEEEtran}
\usepackage{amsfonts}
\usepackage{mathrsfs}
\usepackage{amsfonts}
\usepackage{amssymb}
\usepackage{graphicx}
\usepackage{amsmath}
\usepackage{array}
\usepackage{cases}
\usepackage{algorithm2e}
\usepackage{subfigure}
\usepackage{cite,graphicx,amsmath,amssymb,color}

\usepackage[margin=2.5cm]{geometry}
\usepackage{multirow}

\begin{document}

\title{Convergence-Optimal Quantizer Design of Distributed Contraction-based Iterative Algorithms with Quantized Message Passing}

\newtheorem{Thm}{Theorem}
\newtheorem{Lem}{Lemma}
\newtheorem{Cor}{Corollary}
\newtheorem{Def}{Definition}
\newtheorem{Exam}{Example}
\newtheorem{Alg}{Algorithm}
\newtheorem{Prob}{Problem}
\newtheorem{Rem}{Remark}
\newtheorem{Exm}{Example}

\author{Ying~Cui, \IEEEmembership{Student~Member,~IEEE}, Vincent K.~N.~Lau, \IEEEmembership{Senior~Member,~IEEE},
\\
Department of ECE, The Hong Kong University of Science and Technology\\
{\em cuiying@ust.hk}, {\em eeknlau@ee.ust.hk}}

\maketitle

\begin{abstract}
In this paper, we study the convergence behavior of distributed
iterative algorithms  with quantized message passing. We first
introduce general iterative function evaluation algorithms for
solving fixed point problems distributively. We then analyze the
convergence of the distributed algorithms, e.g. Jacobi scheme and
Gauss-Seidel scheme, under the quantized message passing. Based on
the closed-form convergence performance derived, we propose two
quantizer designs, namely the {\em time invariant
convergence-optimal quantizer} (TICOQ) and the {\em time varying
convergence-optimal quantizer} (TVCOQ),  to minimize the effect of
the quantization error on the convergence. We also study the
tradeoff between the convergence error and message passing overhead
for both TICOQ and TVCOQ. As an example, we apply the TICOQ and
TVCOQ designs to the iterative waterfilling algorithm of MIMO
interference game.
\end{abstract}

\section{Introduction}

Distributed algorithm design and analysis is a very important topic
with important applications in many areas such as deterministic
network utility maximization (NUM) for wireless networks and
non-cooperative game. For example, in \cite{RS:52,RS:53}, the
authors derived various distributed algorithms for a generic
deterministic NUM problem using the decomposition techniques, which
can be classified into primal decomposition and dual decomposition
methods. In \cite{Yudistributedmultiuser:2002}, the authors
investigated a distributed power control algorithm for an
interference channel using non-cooperative game and derived an {\em
iterative water-filling} algorithm to approach the Nash equilibrium
(NE). The interference game problem was extended to iterative
waterfilling algorithm for a wideband interference game with
time/frequency offset in \cite{PalomarSynchronizationerror:2007} and
an iterative precoder optimization algorithm for a MIMO interference
game in \cite{PalomarmuMIMOunifiedview:2008,PalomarmuMIMOwf:2009}.
The authors established a unified convergence proof of the iterative
water-filling algorithms for the SISO frequency-selective
interference game and the MIMO interference game using a contraction
mapping approach. Using this framework, the iterative best response
update (such as the iterative power water-filling as well as the
iterative precoder design) can be regarded as an {\em iterative
function evaluations} w.r.t. a certain contraction mapping and the
convergence property can be easily established using {\em fixed
point
theory}\cite{Bertsekasbookparalleldistributed:1989,fixedpttheory:2001}.
In all these examples, the {\em iterative function evaluation
algorithms} involved explicit message passing between nodes in the
wireless networks during the iteration process. Furthermore, these
existing results have assumed perfect message passing during the
iterations.

In practice, explicit message passing during the iterations in the
distributed algorithms requires explicit signaling in wireless
networks. As such, the message passing cannot be perfect and in many
cases, the messages to pass have to be quantized. As a result, it is
very important and interesting to study about the impact of
quantized message passing on the convergence of the distributed
algorithms. Existing studies on the distributed algorithms under
quantized message passing can be classified into two categories,
namely the {\em distributed quantized average consensus algorithms}
\cite{Kashyapquantizedconsensus:2006,Rabbatquantizedavrconsensus:2007,ConsensusCarli:2008,ConsensusKar:2010,ConsensusNedic:2009,ConsensusYildiz:2007}
as well as the {\em distributed quantized incremental subgradient
algorithms}\cite{Rabbatquantizedincalg:2005,IncSubNedic:2007,IncSubgNedic:2008,IncSubNedicdetnoise:2009}.
For the distributed quantized average consensus algorithms, existing
works considered the algorithm convergence performance under
quantized message passing for uniform quantizer
\cite{Kashyapquantizedconsensus:2006,Rabbatquantizedavrconsensus:2007,ConsensusKar:2010,ConsensusNedic:2009,ConsensusYildiz:2007}
and logarithmic quantizer \cite{ConsensusCarli:2008} with fixed
quantization rate. In \cite{ConsensusKar:2010,ConsensusYildiz:2007},
the authors also considered  quantization interval optimization (for
average consensus algorithms) based on the uniform fixed-rate
quantization structure. Similarly, for the second category of
quantized incremental subgradient algorithms, the authors in
\cite{Rabbatquantizedincalg:2005,IncSubNedic:2007,IncSubgNedic:2008,IncSubNedicdetnoise:2009}
considered the convergence performance of fixed-rate uniform
quantization. In this paper, we are interested in the convergence
behavior of distributed iterative algorithms for solving general
fixed point problems under quantized message passing. The above
works on quantized message passing cannot be applied to our case due
to the following reasons. First of all, the algorithm dynamics of
the existing works (linear dynamics for average consensus algorthms
and {\em step-size based} algorithms for incremental subgradient
algorithms) are very different from the {\em contraction-based
iterative} algorithms we are interested in (for solving fixed point
problems). Secondly, the above works have imposed simplifying
constraints of {\em uniform} and {\em fixed rate} quantizer design
and it is not known if a more general quantizer design or adaptive
quantization rate could further improve the convergence performance
of the iterative algorithms. There are a few technical challenges
regarding the study of convergence behavior in distributed
contraction-based iterative function evaluations.
\begin{itemize}
\item \textbf{Convergence Analysis and Performance Tradeoff under Quantized Message Passing}:
In the literature, convergence of distributed iterative function
evaluation algorithms under quantized message passing has not been
considered. The general model under quantized message passing and
how does the quantization error affect the convergence are not fully
studied. Furthermore, it will also be interesting to study the
tradeoff between convergence error and message passing overhead.
\item \textbf{Quantizer Design based on the Convergence Performance}:
Given the convergence analysis results, how to optimize the
quantizer to minimize the effect of the quantization error on the
convergence is a difficult problem. In general, quantizers are
designed w.r.t. a certain {\em distortion measure} such as the mean
square error \cite{GrayQuantization:1998,GrayQuantizationBook:1992}.
However, it is not clear which distortion measure we should use to
design the quantizer in order to optimize the convergence
performance of the iterative algorithms we considered. Furthermore,
the convergence performance  highly depends on the quantizer
structure as well as the quantization rate, and hence, a
low-complexity solution to the nonlinear integer quantizer
optimization problem is of great importance.
\end{itemize}

In this paper, we shall attempt to shed some lights on these
questions. We shall first introduce a general iterative function
evaluation algorithm with distributed message passing for solving
fixed point problems. We shall then analyze the convergence of the
distributed algorithms, e.g. Jacobi scheme and Gauss-Seidel scheme,
under the quantized message passing. Based on the analysis, we shall
propose two rate-adaptive quantizer designs, namely the {\em time
invariant convergence-optimal quantizer} (TICOQ) and the {\em time
varying convergence-optimal quantizer} (TVCOQ), to minimize the
effect of the quantization error on the convergence. We shall also
develop efficient algorithms to solve the nonlinear integer
programming problem associated with the quantizer optimization
problem. As an illustrative example, we shall apply the TICOQ and
TVCOQ designs to the iterative waterfilling algorithm of the MIMO
interference
game\cite{PalomarmuMIMOunifiedview:2008,PalomarmuMIMOwf:2009}.

%

We first list the important notations in this paper in table
\ref{table:notation}.
\begin{table}[h]
\begin{center} {\footnotesize
\begin{tabular}{|c|c|}
\hline
$n$ & dimension of vector of state variables \\ $m$ ($1\leq m \leq n$) & element index of vector  \\
$K$ & number of nodes/blocks \\
$k$ ($1\leq k \leq K$) & node index/block index\\
$\bar T$ &   total number of iterations  \\ $t$ ($1\leq t \leq \bar T$) & iteration index \\
$\mathcal Q_k$  & component quantizer of node $k$ (general)  \\  $\boldsymbol{\mathcal{Q}}=(\mathcal Q_1,\cdots,\mathcal Q_K)$ & system quantizer (general)\\
superscript $s$ & scalar quantizer (SQ)   \\ superscript $v$ & vector quantizer (VQ)\\
$\mathcal Q^s_k=(\mathcal Q^s_m)_{m \in \mathcal M_k}$ & component quantizer of node $k$ (SQ) \\ $\boldsymbol{\mathcal{Q}}^s=(\mathcal Q^s_1,\cdots,\mathcal Q^s_n)$ & system quantizer (SQ)\\
$\mathbf I^s=(I^s_1,\cdots,I^s_n)$ & quantization index vector (SQ) \\ $\mathbf L^s=(L^s_1,\cdots,L^s_n)$ & quantization rate vector (SQ)\\
$\mathcal Q^v_k$ & component quantizer of node $k$ (VQ) \\ $\boldsymbol{\mathcal{Q}}^v=(\mathcal Q^v_1,\cdots,\mathcal Q^v_K)$ & system quantizer (VQ)\\
$\mathbf I^v=(I^v_1,\cdots,I^v_K)$ & quantization index vector (VQ) \\ $\mathbf L^v=(L^v_1,\cdots,L^v_K)$ & quantization rate vector (VQ)\\
$\mathbb R^+$ & the set of nonnegative real numbers \\ $\mathbb Z^+$
&
the set of nonnegative integers\\
 \hline
\end{tabular}}
\end{center}
\caption{\footnotesize List of Important Notations.}
\label{table:notation}
\end{table}


\section{Iterative Function Evaluations}
In this section, we shall introduce the basic iterative function
evaluation algorithm to solve fixed point problems as well as its
parallel and distributed implementations. We shall then review the
convergence property under perfect message passing in the iteration
process. We shall also illustrate the application of the framework
using the MIMO interference game in
\cite{PalomarmuMIMOunifiedview:2008,PalomarmuMIMOwf:2009} as an
example.
\subsection{A General Framework of Iterative Function Evaluation Algorithms}
In algorithm designs of wireless systems, many iterative algorithms
can be described as the following dynamic update
equation\cite{Bertsekasbookparalleldistributed:1989}:
\begin{align}
\mathbf{x}(t+1)=\mathbf{T}\big(\mathbf{x}(t)\big)\label{eqn:contr-update}
\end{align}
where $\mathbf{x}(t)\in \mathbb{R}^n$ is the vector of state
variables of the system at (discrete) time $t$ and $\mathbf{T}$ is a
mapping from a subset $\boldsymbol{\boldsymbol{\mathcal{X}}}
\subseteq \mathbb{R}^n$ into itself. Such iterative algorithm with
dynamics described by \eqref{eqn:contr-update} is called the {\em
iterative function evaluation algorithm}, which is widely used to
solve {\em fixed point
problems}\cite{fixedpttheory:2001,Bertsekasbookparalleldistributed:1989}.
Specifically, any vector $\mathbf{x}^*\in \boldsymbol{\mathcal{X}}$
satisfying $\mathbf{T}(\mathbf{x}^*)=\mathbf{x}^*$ is called a {\em
fixed point} of $\mathbf{T}$. If the sequence $\{\mathbf{x}(t)\}$
converges to some $\mathbf{x}^*\in \boldsymbol{\mathcal{X}}$ and
$\mathbf{T}$ is continuous at $\mathbf{x}^*$, then $\mathbf{x}^*$ is
a fixed point of $\mathbf{T}$
\cite{Bertsekasbookparalleldistributed:1989}. Therefore, the
iteration in \eqref{eqn:contr-update} can be viewed as an algorithm
for finding such a fixed point of $\mathbf{T}$. We shall first
review a few properties below related to the convergence of
\eqref{eqn:contr-update}. Specifically, $\mathbf{T}$ is called a
contraction mapping if it satisfies some property, which is defined
as follows:
\begin{Def}[Contraction Mapping]
Let
$\mathbf{T}:\boldsymbol{\mathcal{X}}\rightarrow\boldsymbol{\mathcal{X}}$
be a mapping from a subset $\boldsymbol{\mathcal{X}} \subseteq
\mathbb{R}^n$ into itself satisfying the following property
$\|\mathbf{T}(\mathbf{x})-\mathbf{T}(\mathbf{y})\|\leq \alpha
\|\mathbf{x}-\mathbf{y}\|$ ($\forall \mathbf x, \mathbf y \in
\boldsymbol{\mathcal{X}}$),
where $\|\cdot\|$ is some norm and $\alpha\in[0,1)$ is a constant
scalar. Then the mapping $\mathbf{T}$ is called a {\em contraction
mapping} and the scalar $\alpha$ is called the {\em modulus} of
$\mathbf{T}$.~ \hfill\QED\label{def:contr-mapping}
\end{Def}

\begin{Rem}(\emph{Comparison with Step-size Based Incremental Subgradient
Algorithms}) The incremental subgradient
algorithms\cite{Bertsekasincsubgradient:1999}  can be described as
$\mathbf{x}(t+1)=\mathbf{x}(t)-\epsilon_t g\big(\mathbf{x}(t)\big)$,
where $\{\epsilon_t\}$ is the step-size sequence and
$g\big(\mathbf{x}(t)\big)$ is a subgradient of the objective
function at $\mathbf{x}(t)$ in a minimization problem. Such
step-size based update algorithms and their associated convergence
dynamics are quite different from the iterative function evaluation
algorithm we considered in \eqref{eqn:contr-update}. ~ \hfill\QED
\end{Rem}

If $\mathbf{T}$ is a contraction mapping, then the iterative update
in \eqref{eqn:contr-update} is called {\em contracting iteration}.
The convergence of \eqref{eqn:contr-update} is summarized as follows
(Proof can be found in
\cite{Bertsekasbookparalleldistributed:1989}):
\begin{Thm}[Convergence of Contracting Iterations]
Suppose that
$\mathbf{T}:\boldsymbol{\mathcal{X}}\rightarrow\boldsymbol{\mathcal{X}}$
is a contraction mapping with modulus $\alpha\in[0,1)$ and that
$\boldsymbol{\mathcal{X}} \subseteq \mathbb{R}^n$ is closed. We
have:

(1) (Existence and Uniqueness of Fixed Points) The mapping
$\mathbf{T}$ has a unique fixed point $\mathbf{x}^{*}\in
\boldsymbol{\mathcal{X}}$.

(2) (Geometric Convergence) For any initial vector $\mathbf{x}(0)\in
\boldsymbol{\mathcal{X}}$, the sequence $\{\mathbf{x}(t)\}$
generated by \eqref{eqn:contr-update} converges to $\mathbf{x}^{*}$
geometrically. In particular, $\|\mathbf{x}(t)-\mathbf{x}^{*}\|\leq
\alpha^t \|\mathbf{x}(0)-\mathbf{x}^*\|\quad \forall t\geq 0$.
~ \hfill\QED \label{Thm:contr-mapping-cov}
\end{Thm}

In the above discussion, $\|\cdot\|$ can be any well-defined norm.
There are many useful norms in the literature. However, the commonly
used norms can be classified into two groups, namely weighted
maximum norm and $L_p$ norm ($1\leq p < \infty$). They are
elaborated below:
\begin{itemize}
\item Weighted maximum norm:
\begin{align}
\|\mathbf{x}\|^{\mathbf{a}}_{\infty}=\max_m \frac{|x_m|}{a_m}\
(a_m>0)  \label{eqn:max-weight-norm-def}
\end{align} Note that for
$a_m=1\ \forall m$, this reduces to the maximum norm, which can also
be obtained from the $L_p$ norm by taking the limit $p\rightarrow
\infty$.

\item $L_p$ norm ($1\leq p <
\infty$):
\begin{align}
 \|\mathbf{x}\|_{p}=\Big(\sum_{m=1}^n
|x_m|^p\Big)^{\frac{1}{p}} \label{eqn:L_p-norm-def}
\end{align}
Note that for $p = 1$ we get the taxicab norm and for $p = 2$ we get
the Euclidean norm.

\end{itemize}

\subsection{Parallel and Distributed Implementation of Contracting Iterations}
In practice, large scale computation always involves a number of
processors or communication nodes jointly executing a computational
task. As a result, parallel and distributed implementation is of
prime importance. Information acquisition and control are within
geographically distributed nodes, in which distributed computation
is more preferable. In this part, we shall discuss the efficient
parallel and distributed computation of the contracting iteration in
\eqref{eqn:contr-update}.

To perform efficient parallel and distributed implementations with
$K$ processors, the set $\boldsymbol{\mathcal{X}}$ is partitioned
into a Cartesian product of lower dimensional sets, based on the
computational complexity consideration or the local information
extraction and control requirement. Mathematically, it can be
expressed as $\boldsymbol{\mathcal{X}}=\prod_{k=1}^K
\boldsymbol{\mathcal{X}}_k$, where $\boldsymbol{\mathcal{X}}_k
\subseteq \mathbb{R}^{n_k}$ and $\sum_{k=1}^K n_k=n$. Let $n_0=0$
and $\mathcal M_k=\{m\in \mathbb N: \sum_{l=1}^k n_{l-1} +1 \leq m
\leq \sum_{l=1}^k n_l\}$ be the index set of the $k$-th component
set $\boldsymbol{\mathcal X}_k$ ($1\leq k \leq K$), where $\mathbb
N$ is the set of integers. Thus, $\boldsymbol{\mathcal
X}_k=\prod_{m\in \mathcal M_k} \mathcal X_m$, where $\mathcal X_m
\subseteq \mathbb R^1$. Any vector $\mathbf{x}\in
\boldsymbol{\mathcal{X}}$ is decomposed as
$\mathbf{x}=(\mathbf{x}_1,\cdots,\mathbf{x}_K)$ with the $k$-th
block component $\mathbf{x}_k=(x_m)_{m\in \mathcal M_k} \in
\boldsymbol{\mathcal{X}}_k=\prod_{m\in \mathcal M_k} \mathcal X_m$
and the mapping
$\mathbf{T}:\boldsymbol{\mathcal{X}}\rightarrow\boldsymbol{\mathcal{X}}$
is decomposed as
$\mathbf{T}(\mathbf{x})=\big(\mathbf{T}_1(\mathbf{x}),\cdots,
\mathbf{T}_K(\mathbf{x})\big)$ with the $k$-th block component
$\mathbf{T}_k:\boldsymbol{\mathcal{X}}\rightarrow\boldsymbol{\mathcal{X}}_k$.


When the set $\boldsymbol{\mathcal{X}}$ is a Cartesian product of
lower dimensional sets $\boldsymbol{\mathcal{X}}_k$,
block-parallelization with $K$ processors can be implemented by
assigning one processor to update a different block component. The
most common updating strategies for
$\mathbf{x}_1,\cdots,\mathbf{x}_K$ based on the block mapping
$\mathbf{T}$ are:
\begin{itemize}
\item \textbf{Jacobi Scheme}: All block components
$\mathbf{x}_1,\cdots,\mathbf{x}_K$ are updated simultaneously, i.e.
\begin{align}
\mathbf{x}_k(t+1)=\mathbf{T}_k\big(\mathbf{x}(t)\big),\quad 1\leq
k\leq K\label{eqn:update-component}
\end{align}

\item \textbf{Gauss-Seidel Scheme}: All block components
$\mathbf{x}_1,\cdots,\mathbf{x}_K$ are updated sequentially, one
after the other, i.e.
\begin{align}
\mathbf{x}_k(t+1)=\mathbf{S}_k\big( \mathbf{x}(t) \big),\quad 1\leq
k\leq K \label{G-S-iteration}
\end{align}
where
$\mathbf{S}_k:\boldsymbol{\mathcal{X}}\rightarrow\boldsymbol{\mathcal{X}}_k$
given by
\begin{align}
&\mathbf{S}_k(\mathbf{x})\label{G-S-mapping}\\
=&\left\{
\begin{array}{ll}\mathbf{T}_k(\mathbf{x}), & k=1\\
\mathbf{T}_k\big(\mathbf{S}_1(\mathbf{x}),\cdots,\mathbf{S}_{k-1}
(\mathbf{x}),\mathbf{x}_k,\cdots,\mathbf{x}_K\big), & 2\leq k \leq K
\end{array} \right.\nonumber
\end{align}
is the $k$-th block component of the Guass-Seidel mapping $\mathbf
S:\boldsymbol{\mathcal{X}}\rightarrow\boldsymbol{\mathcal{X}}$, i.e.
$\mathbf{S}(\mathbf{x})=\big(\mathbf{S}_1(\mathbf{x}),\cdots,
\mathbf{S}_K(\mathbf{x})\big)$.


\end{itemize}
Both  Jacobi Scheme and Gauss-Seidel Scheme belong to synchronous
update schemes\footnote{Due to page limit, we shall illustrate the
design for synchronous updates in \eqref{eqn:update-component} and
\eqref{G-S-iteration}. However, the scheme can be extended to deal
with {\em totally asynchronous updates} easily
\cite{Bertsekasbookparalleldistributed:1989}, which will be further
illustrated later in footnote \ref{ft:asyn}.}. Specifically, Jacobi
Scheme assumes the network is synchronized, while the Gauss-Seidel
Scheme assumes the network provides a (Hamiltonian) cyclic
route\cite{Bertsekasbookparalleldistributed:1989}.

The general weighted block-maximum norm on $\mathbb{R}^{n}$, which
is usually associated with the block partition of the vector
$\mathbf{x}$,  is defined
as\cite{Bertsekasbookparalleldistributed:1989}:
\begin{align}
\|\mathbf{x}\|^{\mathbf{w}}_{\text{block}}=\max_k
\frac{\|\mathbf{x}_k\|_k}{w_k}\label{eqn:weighted-block-max-norm}
\end{align}
where $\mathbf{w}=(w_1,\cdots, w_K)>0$ is the vector weight and
$\|\cdot\|_k$ is the norm for the $k$-th block
component\footnote{Since in general, the norm of each block
component may not be the same, subscript $k$ is used in
$\|\cdot\|_k$. } $\mathbf x_k$, which can be any given norm on
$\mathbb{R}^{n_k}$, such as the weighted maximum norm and the $L_p$
norm ($1\leq p < \infty$) defined in \eqref{eqn:max-weight-norm-def}
and \eqref{eqn:L_p-norm-def}.
The mapping
$\mathbf{T}:\boldsymbol{\mathcal{X}}\rightarrow\boldsymbol{\mathcal{X}}$
is called a {\em block-contraction} with modulus $\alpha\in[0,1)$ if
it is a contraction under the above induced weighted block-maximum
norm $\|\cdot\|^{\mathbf{w}}_{\text{block}}$ with modulus $\alpha$.
The convergence of the Jacobi scheme and Gauss-Seidel scheme based
on the block-contraction is summarized in the following theorem
\cite{Bertsekasbookparalleldistributed:1989}:
\begin{Thm}(\emph{Convergence of Jacobi Scheme and Gauss-Seidel Scheme}) If
$\mathbf{T}:\boldsymbol{\mathcal{X}}\rightarrow\boldsymbol{\mathcal{X}}$
is a block-contraction, then the Gauss-Seidel mapping $\mathbf{S}$
is also a block-contraction with the same modulus as $\mathbf{T}$.
Furthermore, if $\boldsymbol{\mathcal{X}}$ is closed, then the
sequence $\{\mathbf{x}(t)\}$ generated by both the Jocobi scheme in
\eqref{eqn:update-component} and the Gauss-Seidel scheme in
\eqref{G-S-iteration} based on the mapping $\mathbf{T}$ converges to
the unique fixed point $\mathbf{x}^*\in \boldsymbol{\mathcal{X}}$ of
$\mathbf{T}$ geometrically. ~ \hfill\QED
\label{Thm:Jacobi-Gauss-Seidel-cov}
\end{Thm}

\subsection{Application Example --- MIMO Interference Game}\label{subsec_application-example}
The contracting iteration in \eqref{eqn:contr-update} is very useful
in solving fixed point problems. Fixed point problem is highly
related to distributed resource optimization problems in wireless
systems\cite{Yudistributedmultiuser:2002,Huangthesis:2005,PalomarmuMIMOunifiedview:2008,PalomarmuMIMOwf:2009}.
For example, finding the {\em Nash Equilibrim (NE)} of a game is a
fixed point problem. In this subsection, we shall illustrate the
application of contracting iterations using MIMO interference game
\cite{PalomarmuMIMOunifiedview:2008,PalomarmuMIMOwf:2009} as an
example.

Consider a system with  $K$ noncooperative transmitter-receiver
pairs communicating simultaneously over a MIMO channel with $N$
transmit antenna and $N$ receive
antenna\cite{PalomarmuMIMOunifiedview:2008,PalomarmuMIMOwf:2009}.
The received signal of the $k$-th receiver is given by:
\begin{align}
\mathbf{y}_k= \mathbf H_{kk} \mathbf s_k +\sum_{j\neq k} \mathbf
H_{jk} \mathbf s_j+ \mathbf n_k \label{eqn:signal-model-MIMO-game}
\end{align}
where $\mathbf s_k\in \mathbb{C}^N$ and $\mathbf y_k\in
\mathbb{C}^N$ are the vector transmitted by the $k$-th transmitter
and the vector received by the $k$-th receiver respectively,
$\mathbf H_{kk}\in \mathbb{C}^{N\times N}$ is the direct-channel of
the $k$-th link, $\mathbf H_{jk}\in \mathbb{C}^{N\times N}$ is the
cross-channel from the $j$-th transmitter to the $k$-th receiver,
and $\mathbf n_k \in \mathbb{C}^N$ is a zero-mean circularly
symmetric complex Gaussian noise vector with covariance matrix
$\mathbf R_{\mathbf n_k}$.  For each transmitter $k$, the total
average transmit power is given by
\begin{align}
\mathbb E \big[ \|\mathbf s_k\|^2_2\big]=\text{Tr}(\mathbf P_k)\leq
P_k\label{eqn:pwr-cosnt-MIMO-game}
\end{align}
where $\text{Tr}(\cdot)$ denotes the trace operator, $\mathbf
P_k\triangleq \mathbb E[\mathbf{s}_k \mathbf{s}_k^H]$ is the
covariance matrix of the transmitted vector $\mathbf{s}_k$ and $P_k$
is the maximum average transmitted power. The maximum throughput of
link $k$ for a given set of users' covariance matrices $\mathbf P_1,
\cdots, \mathbf P_K$ is as follows
\begin{align}
r_k (\mathbf P_k, \mathbf P_{-k})=\log \det\big(\mathbf I +
\mathbf{H}_{kk}^{H}\mathbf{R}_{-k}^{-1}(\mathbf
P_{-k})\mathbf{H}_{kk}\mathbf P_k\big)
\label{eqn:throughput-MIMO-game}
\end{align}
where $\mathbf{R}_{-k}(\mathbf
P_{-k})\triangleq\mathbf{R}_{\mathbf{n}_k}+\sum_{j\neq
k}\mathbf{H}_{jk}\mathbf P_j\mathbf{H}_{jk}^H$ is the noise
covariance matrix plus the MUI observed by user $k$, and $\mathbf
P_{-k}\triangleq(\mathbf P_j)_{j\neq k}$ is the covariance matrix of
all other users except user $k$.

In the MIMO interference game
\cite{PalomarmuMIMOunifiedview:2008,PalomarmuMIMOwf:2009}, each
player $k$ competes against the others by choosing his transmit
covariance matrix $\mathbf P_k$ (i.e., his strategy) that maximizes
his own maximum throughput $r_k (\mathbf P_k, \mathbf P_{-k})$ in
\eqref{eqn:throughput-MIMO-game}, subject to the transmit power
constraint in \eqref{eqn:pwr-cosnt-MIMO-game}, the mathematical
structure of which is as follows
\begin{align}
(\mathcal G) \max_{\mathbf P_k} & \quad r_k (\mathbf P_k, \mathbf
P_{-k})
\quad \forall k \label{eqn:MIMO-game-formu}\\
s.t. & \quad \mathbf P_k \in \mathscr{P}_k \nonumber
\end{align}
where $\mathscr{P}_k\triangleq \Big\{\mathbf P_k \in
\mathbb{C}^{N\times N}: \mathbf P_k \succcurlyeq 0, \text{Tr}
(\mathbf P_k)=P_k\Big\}$ is the admissible strategy set of user $k$,
and  $\mathbf P_k \succcurlyeq 0$ denotes that $\mathbf P_k$ is a
positive semidefinite matrix.
Given $k$ and $\mathbf P_{-k}\in \mathscr{P}_{-k}$, the solution to
the non-cooperative game \eqref{eqn:MIMO-game-formu} is the
well-know waterfilling solution $\mathbf P^*_k=\mathbf{WF}_k(\mathbf
P_{-k})$, where the waterfilling operator $\mathbf{WF}_k(\mathbf
P_{-k})$  can be equivalently written as
\cite{PalomarmuMIMOunifiedview:2008}
\begin{align}
\mathbf{WF}_k(\mathbf
P_{-k})=\big[-\big(\mathbf{H}_{kk}^{H}\mathbf{R}_{-k}^{-1}(\mathbf
P_{-k})\mathbf{H}_{kk}\big)^{-1}\big]_{\mathscr{P}_k}
\label{eqn:given-others-solution}
\end{align}
where $\big[\mathbf{X}_0\big]_{\mathscr{X}}=\arg \min_{\mathbf{Z}\in
\mathscr{X}}\|\mathbf{Z}-\mathbf{X}_0\|_F$ denotes the matrix
projection of $\mathbf{X}_0$ w.r.t Frobenius norm\footnote{If we
arrange $MN$ elements of a $M\times N$ matrix $\mathbf X$ as a
$MN$-dimensional vector $\mathbf x$, then the Frobenius norm of
matrix $\mathbf X$ is equivalent to the $L_2$ norm of vector
$\mathbf x$.} $\|\cdot\|_F$ onto the set $\mathscr{X}$. The NE of
the MIMO Gaussian interference game is the fixed point solution of
the waterfilling mapping $\mathbf{WF}:\mathscr{P}\rightarrow
\mathscr{P}$, i.e. $\mathbf P_k^{*}=\mathbf{WF}_k(\mathbf
P_{-k}^{*})\ \forall k $, where $\mathscr{P}\triangleq
\mathscr{P}_1\times \cdots \times \mathscr{P}_K$ and
$\mathbf{WF}=(\mathbf{WF}_1,\cdots,\mathbf{WF}_K)$.

In \cite{PalomarmuMIMOunifiedview:2008}, it is shown that under some
mild condition, the mapping $\mathbf{WF}$ is a
block-contraction\footnote{After rearranging the elements of the
$N\times N$ covariance matrix $\mathbf P_k$ as a $N^2$-dimensional
vector, the block-contraction $\mathbf{WF}$ w.r.t.
$\|\cdot\|^{\mathbf{w}}_{F,\text{block}}$ is equivalent to a
block-contraction w.r.t. $\|\cdot\|^{\mathbf{w}}_{\text{block}}$
defined in \eqref{eqn:weighted-block-max-norm} with each
$\|\cdot\|_k$ being $L_2$ norm.} w.r.t.
$\|\cdot\|^{\mathbf{w}}_{F,\text{block}}$. Therefore, the NE can be
achieved by the following contracting iteration
\begin{align}
\mathbf P(t+1)=\mathbf{WF}\big(\mathbf P(t)\big)
\label{eqn:mimo-centrl-iter}
\end{align}
where $\mathbf P=(\mathbf P_1,\cdots,\mathbf P_K)$. It can be easily
seen that the waterfilling algorithm for the MIMO interference game
in \eqref{eqn:mimo-centrl-iter} is a special case of the contracting
iterations in \eqref{eqn:contr-update}. In our general model,
$\mathbf{x}$ in \eqref{eqn:contr-update} corresponds to $\mathbf P$
in \eqref{eqn:mimo-centrl-iter}; the block-contraction mapping
$\mathbf{T}$ in \eqref{eqn:contr-update} corresponds to
$\mathbf{WF}$ in \eqref{eqn:mimo-centrl-iter}; the $k$-th block
component $\mathbf{x}_k$ corresponds to the covariance matrix
$\mathbf P_k$; the $k$-th block component mapping $\mathbf{T}_k$
corresponds to $\mathbf{WF}_k$.

For the parallel and distributed implementation, we can partition
the variable space $\mathscr{P}= \prod_k \mathscr{P}_k$, where each
variable space $\mathscr{P}_k$ corresponds to the covariance matrix
of the $k$-th link. In each iteration, the receiver of each link $k$
needs to locally measure the PSD of the interference received from
the transmitter of the other links, i.e. $\sum_{j\neq
k}\mathbf{H}_{jk}\mathbf P_j(t)\mathbf{H}_{jk}^H$, computes the
covariance matrix of the $k$-th link and transmits the computational
results to the associated transmitter. There are two distributed
iterative waterfilling algorithms (IWFA) based on this waterfilling
block-contraction, namely simultaneous IWFA and sequential IWFA,
which are described as follows:
\begin{itemize}
\item \textbf{Simultaneous IWFA}: It is an example of the Jacobi
scheme, which is given by
\[\mathbf P_k(t+1)=\mathbf{WF}\big(\mathbf P_{-k}(t)\big), \ 1 \leq k \leq K
\]
\item \textbf{Sequential IWFA}: It is an example of the Gauss-Seidel scheme,
which is given by
\[
\mathbf P_k(t+1)=\left\{
\begin{array}{ll} \mathbf{WF}\big(\mathbf P_{-k}(t)\big), &
\text{if}
(t+1) \text{mod} K=k
\\
\mathbf P_k(t), & \textrm{otherwise}
\end{array} \right.
\]
\end{itemize}

%
%
%

\section{Contracting Iterations under Quantized Message
Passing}\label{Sec:q-message-passing-model}

In this section, we shall study the impact of the quantized message
passing on the contracting iterations. We shall first introduce a
general quantized message passing model, followed by some general
results regarding the convergence behavior under quantized message
passing.

\subsection{General Model of Quantized Message Passing}
\label{Subsec:general-model-Q-message-passing}

%

\begin{figure}[h]
\begin{center}
\includegraphics[height=4cm, width=8cm]{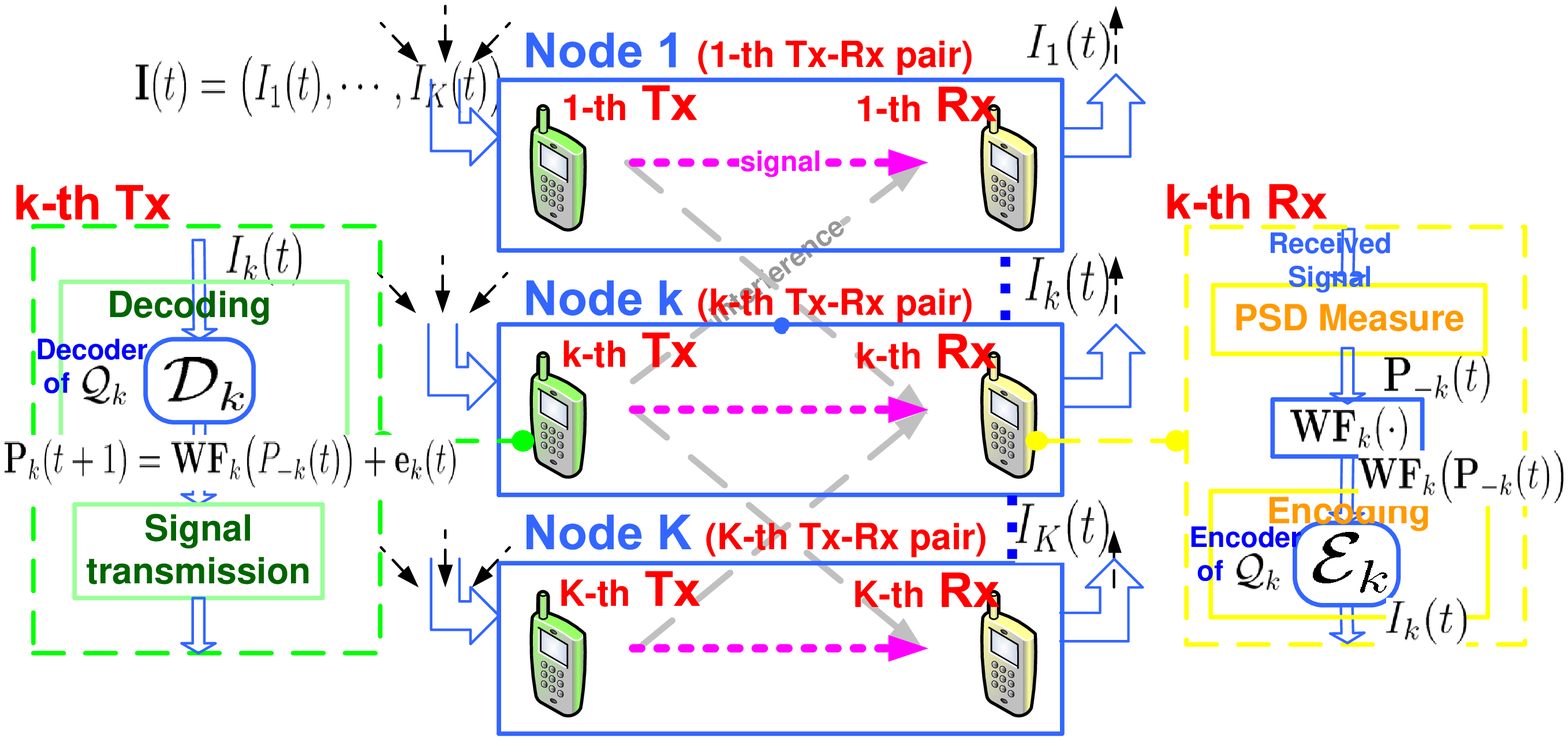}
\caption{Illustration of $K$-pair MIMO interference game.
}\label{message_passing}
\end{center}
\end{figure}

We assume there are $K$ processing nodes geographically distributed
in the wireless systems. Fig. \ref{message_passing} illustrates an
example of $K$-pair MIMO interference game with quantized message
passing.  The system quantizer can be characterized by the tuple
$\boldsymbol{\mathcal Q}=(\mathcal Q_1,...,\mathcal Q_K)$, where
$\mathcal Q_k$ is the component quantizer (can be scalar or vector
quantizer) for the $k$-th node. $\mathcal Q_k$ can be further
denoted by the tuple $\mathcal Q_k=(\mathcal E_k, \mathcal D_k)$.
$\mathcal E_k:\boldsymbol{\mathcal{X}}_k\rightarrow \mathcal{I}_k $
is an {\em encoder} and $\mathcal D_k: \mathcal{I}_k \rightarrow
\hat{\boldsymbol{\mathcal{X}}}_k $ is a {\em decoder}. $\mathcal I_k
= \{1,\cdots,2^{L_k}\}$ and $L_k$ are the {\em index set} and the
{\em quantization rate} of the component quantizer $\mathcal Q_k$.
$\hat{\boldsymbol{\mathcal{X}}}_k$ is the {\em reproduction
codebook}, which is the set of all possible quantized outputs of
$\mathcal Q_k$\cite{GrayQuantization:1998}. The quantization rule is
completely specified by $\mathcal
Q_k:\boldsymbol{\mathcal{X}}_k\rightarrow
\hat{\boldsymbol{\mathcal{X}}}_k$. Specifically, the quantized value
is given by $\hat{\mathbf x}_k=\mathcal Q_k (\mathbf x_k)=\mathcal
D_k \big(\mathcal E_k (\mathbf x_k)\big)$. Each node $k$ updates the
$k$-th block component $\mathbf{x}_k$ of the $n$-dimensional vector
$\mathbf{x}$, i.e. computes
$\mathbf{x}_k(t+1)=\mathbf{T}_k\big(\mathbf{x}(t)\big)$. The encoder
$\mathcal E_k$ of $\mathcal Q_k$ accepts the input
$\mathbf{T}_k\big(\mathbf{x}(t)\big)$ and produces a quantization
index $I_k(t)=\mathcal
E_k\big(\mathbf{T}_k\big(\mathbf{x}(t)\big)\big)$. Each node $k$
broadcasts the quantization index $I_k(t)$. In other words, the
message passing involves only the quantization indices $\mathbf
I(t)=\big(I_k(t),\cdots,I_K(t)\big)$ instead of the actual controls
$\mathbf{T}\big(\mathbf{x}(t)\big)=\big(\mathbf{T}_1\big(\mathbf{x}(t)\big),\cdots,\mathbf{T}_K\big(\mathbf{x}(t)\big)\big)$.
Upon receiving the quantization index $I_k(t)$, the decoder
$\mathcal D_k$ of $\mathcal Q_k$ produces a quantized value
$\mathbf{x}_k(t+1)=\hat{\mathbf{T}}_k\big(\mathbf{x}(t)\big)=\mathcal
D_k\big(I_k(t)\big)=\mathbf{T}_k\big(\mathbf{x}(t)\big)+\mathbf{e}_k(t)$.
Therefore, the contracting iteration update dynamics of
\eqref{eqn:contr-update} with quantized message passing can be
modified as:
\begin{align}
\mathbf{x}(t+1)=\mathbf{T}\big(\mathbf{x}(t)\big)+\mathbf{e}(t)\label{eqn:contr-update-error}
\end{align}
where $\mathbf{e}(t)\in \mathbb{R}^n$ is the {\em quantization error
vector} at time $t$. The quantizer design affects the convergence
property of the iterative update algorithm fundamentally via the
quantization error random process $\mathbf{e}(t)$. Generally, the
update of each block component is based on the latest overall
vector, because
$\mathbf{T}_k:\boldsymbol{\mathcal{X}}\rightarrow\boldsymbol{\mathcal{X}}_k$.
Thus, the decoders of the system quantizer $\boldsymbol{\mathcal
D}=(\mathcal D_1,\cdots,\mathcal D_K)$ is needed at each node. On
the other hand, the $k$-th node only requires the encoder $\mathcal
E_k$ of the corresponding quantizer component $\mathcal{Q}_k$.

Consider the application example in Section
\ref{subsec_application-example} under quantized message passing.
The system quantizer $\boldsymbol{\mathcal Q}=(\mathcal
Q_1,...,\mathcal Q_K)$ can be applied in  the MIMO interference game
with $K$ noncooperative transmitter-receiver pairs  as illustrated
in Fig. \ref{message_passing}. Specifically, for the $k$-th link,
the encoder $\mathcal E_k$ is placed at receiver and the decoder
$\mathcal D_k$ is placed at the transmitter. The MIMO interference
game under quantized message passing will be illustrated in the
following example:

\begin{Exm}(\emph{MIMO Interference under Quantized Message Passing})
In the $t$-th iteration, the receiver of the $k$-th link locally
measures PSD of the interference received from the transmitter of
the other links, i.e. $\sum_{j\neq k}\mathbf{H}_{jk}\mathbf
P_j(t)\mathbf{H}_{jk}^H$, and computes $\mathbf{WF}_k\big(\mathbf
P_{-k}(t)\big)$. The encoder $\mathcal E_k$ of $\mathcal Q_k$ at the
$k$-th receiver encodes $\mathbf{WF}_k\big(\mathbf P_{-k}(t)\big)$
and passes the quantization index $I_k(t)=\mathcal
E_k\big(\mathbf{WF}_k\big(\mathbf P_{-k}(t)\big)\big)$ to the $k$-th
transmitter. The decoder $\mathcal D_k$ of $\mathcal Q_k$ at $k$-th
transmitter produces a quantized value $\mathbf{P}_k(t+1)=\mathcal
D_k \big(I_k(t)\big) =\mathbf{WF}_k\big(\mathbf
P_{-k}(t)\big)+\mathbf{e}_k(t)$. The contracting iterative update
dynamics of \eqref{eqn:mimo-centrl-iter} for the MIMO interference
game under quantized message passing is given by:
\begin{align}
\mathbf P(t+1)=&\mathbf{WF}\big(\mathbf P(t)\big)+\mathbf{e}(t)
\label{eqn:mimo-centrl-iter-error}
\end{align}
\end{Exm}

\subsection{Convergence Property under Quantized Message Passing}

Under the quantized message passing, the convergence of the
contracting iterations is summarized in the following lemma:
\begin{Lem}
(\emph{Convergence of Contracting Iterations under Quantized Message
Passing}) Suppose that
$\mathbf{T}:\boldsymbol{\mathcal{X}}\rightarrow\boldsymbol{\mathcal{X}}$
is a contraction mapping with modulus $\alpha\in[0,1)$ and fixed
point $\mathbf{x}^{*}\in \boldsymbol{\mathcal{X}}$, and that
$\boldsymbol{\mathcal{X}} \subseteq \mathbb{R}^n$ is closed. For any
initial vector $\mathbf{x}(0)\in \boldsymbol{\mathcal{X}}$, the
sequence $\{\mathbf{x}(t)\}$ generated by
\eqref{eqn:contr-update-error} satisfies:

(a) $\|\mathbf{x}(t)-\mathbf{x}^{*}\|\leq \alpha^t
\|\mathbf{x}(0)-\mathbf{x}^*\|+E(t)$ $\forall t\geq 1$,
where
$E(t)=\alpha^{t-1}\sum_{l=0}^{t-1}\alpha^{-l}\|\mathbf{e}(l)\|$ is
the accumulated error up to the time $t$ induced by the quantized
message passing.

(b) 
For each $t$, if there exists a vector
$\tilde{\mathbf{e}}_t\in\mathbb{R}^n$ such that
$\|\mathbf{e}(t)\|\leq \|\tilde{\mathbf{e}}_t\|$, then $E(t)\leq
\tilde E (t)$, where $\tilde E (t)\triangleq
\alpha^{t-1}\sum_{l=0}^{t-1}\alpha^{-l}\|\tilde{\mathbf{e}}_l\|$.

(c) 
If
$\|\tilde{\mathbf{e}}_1\|=\cdots=\|\tilde{\mathbf{e}}_t\|\triangleq
\|\bar{\mathbf{e}}\|$, then $E(t)\leq \bar E (t)$, where $\bar E
(t)\triangleq \frac{1-\alpha^t}{1-\alpha}\|\bar{\mathbf{e}}\|$ with
limiting error bound  $\bar E(\infty)\triangleq \lim_{t\rightarrow
\infty}\bar E(t)=\frac{\|\bar{\mathbf{e}}\|}{1-\alpha}$.
Furthermore, define the stationary set as $\mathbb S \triangleq
\{\boldsymbol{\mathcal Q}(\mathbf x)
:\|\mathbf{x}-\mathbf{x}^{*}\|\leq \bar E(\infty)\}$. The sufficient
condition for convergence is $\mathbf x= \boldsymbol{ \mathcal Q}
\big( \mathbf T (\mathbf x)\big)$ $\forall \mathbf x \in \mathbb S$
and the necessary condition for convergence is $\exists \mathbf x
\in \mathbb S$, such that $\mathbf x= \boldsymbol{ \mathcal Q} \big(
\mathbf T (\mathbf x)\big)$. \label{Lem:contr-mapping-cov-error}
\end{Lem}
\begin{proof}
Please refer to Appendix A for the proof.
\end{proof}

Note that, in the above lemma, the norm $\|\cdot\|$ can be any
general norm. In the following, we shall focus on characterizing the
convergence behavior of the distributed Jacobi and Gauss-Seidel
schemes under quantized message passing with the underlying
contraction mapping $\mathbf T$ defined w.r.t. the weighted
block-maximum norm $\|\cdot\|^{\mathbf{w}}_{\text{block}}$
\cite{Bertsekasbookparalleldistributed:1989,PalomarmuMIMOunifiedview:2008,PalomarmuMIMOwf:2009}.
Under quantized message passing, the {\em algorithm dynamics} of the
two commonly used parallel and distributed schemes can be described
as follows:

\begin{itemize}
\item \textbf{Jacobi Scheme under Quantized Message Passing}:
\begin{align}
\mathbf{x}_k(t+1)=\mathbf{T}_k\big(\mathbf{x}(t)\big)+\mathbf{e}_k(t),\
1\leq k\leq K\label{eqn:update-component-error}
\end{align}

\item \textbf{Gauss-Seidel Scheme under Quantized Message Passing}:
\begin{align}
\mathbf{x}_k(t+1)=\hat{\mathbf{S}}_k\big( \mathbf{x}(t)
\big)+\mathbf{e}_k(t),\ 1\leq k\leq K
\label{eqn:G-S-iteration-error}
\end{align}
where
\begin{align}
&\hat{\mathbf{S}}_k(\mathbf{x})\label{G-S-mapping-error}\\
=&\left\{
\begin{array}{ll}\mathbf{T}_k(\mathbf{x}), & k=1\\
\mathbf{T}_k\big(\hat{\mathbf{S}}_1(\mathbf{x})+\mathbf{e}_1,\cdots,&\\
\quad \hat{\mathbf{S}}_{k-1} (\mathbf{x})+\mathbf{e}_{k-1},
\mathbf{x}_k,\cdots,\mathbf{x}_K\big), & 2\leq k \leq K
\end{array} \right.\nonumber
\end{align}
\end{itemize}

Applying the results of Lemma \ref{Lem:contr-mapping-cov-error}, the
convergence property of the distributed Jacobi and Gauss-Seidel
schemes in \eqref{eqn:update-component-error} and
\eqref{eqn:G-S-iteration-error} can be summarized in the following
lemma.
\begin{Lem}(\emph{Convergence of Jacobi Scheme and Gauss-Seidel Scheme  under Quantized Message Passing})
Suppose that
$\mathbf{T}:\boldsymbol{\mathcal{X}}\rightarrow\boldsymbol{\mathcal{X}}$
is a block-contraction mapping w.r.t. the weighted block-maximum
norm $\|\cdot\|^{\mathbf{w}}_{\text{block}}$  with modulus
$\alpha\in[0,1)$ and fixed point $\mathbf{x}^{*}\in
\boldsymbol{\mathcal{X}}$, and that $\boldsymbol{\mathcal{X}}
\subseteq \mathbb{R}^n$ is closed. For every initial vector
$\mathbf{x}(0)\in \boldsymbol{\mathcal{X}}$, the sequence
$\{\mathbf{x}(t)\}$ generated by both the Jacobi scheme and the
Gauss-Seidel scheme under quantized message passing in
\eqref{eqn:update-component-error} and
\eqref{eqn:G-S-iteration-error} satisfies\footnote{Our analysis can
be extended to totally asynchronous scheme in which the results of
Lemma \ref{Lem:Jacobi-Gauss-Seidel-cov-error} becomes: (a)
$E^{\mathbf{w}}_{\text{block}}(t)=\frac{1}{1-\alpha}\alpha^{t-1}\sum_{l=0}^{t-1}\alpha^{-l}\|\mathbf{e}(l)\|^{\mathbf{w}}_{\text{block}}$.
(b) $\tilde E^{\mathbf{w}}_{\text{block}}
(t)\triangleq\frac{1}{1-\alpha}
\alpha^{t-1}\sum_{l=0}^{t-1}\alpha^{-l}\|\tilde{\mathbf{e}}_l\|^{\mathbf{w}}_{\text{block}}$.
(c) If $\|\bar{\mathbf
e}\|^{\mathbf{w}}_{\text{block}}<(1-\alpha)\|\mathbf{x}(0)-\mathbf{x}^{*}\|^{\mathbf{w}}_{\text{block}}$,
we have $\bar
E^{\mathbf{w}}_{\text{block}}(t)=\frac{1-\alpha^t}{(1-\alpha)^2}\|\bar{\mathbf{e}}\|^{\mathbf{w}}_{\text{block}}$
and $\bar
E^{\mathbf{w}}_{\text{block}}(\infty)=\frac{1}{(1-\alpha)^2}\|\bar{\mathbf{e}}\|^{\mathbf{w}}_{\text{block}}$.
By the {\em Asynchronous Convergence Theorem} (Proposition 2.1 in
Chapter 6 of \cite{Bertsekasbookparalleldistributed:1989}), we can
prove (c) (similar to the proof of Theorem 12 in
\cite{PalomarmuMIMOwf:2009}). The proof is omitted here due to page
limit. Since the error bound result of totally asynchronous scheme
is similar to Jacobi Scheme and Gauss-Seidel Scheme, our quantizer
design later can be applied to the asynchronous case.
\label{ft:asyn}}

(a)
$\|\mathbf{x}(t)-\mathbf{x}^{*}\|^{\mathbf{w}}_{\text{block}}\leq
\alpha^t
\|\mathbf{x}(0)-\mathbf{x}^*\|^{\mathbf{w}}_{\text{block}}+E^{\mathbf{w}}_{\text{block}}(t)$
$\forall t\geq 1$,
where
$E^{\mathbf{w}}_{\text{block}}(t)=\alpha^{t-1}\sum_{l=0}^{t-1}\alpha^{-l}\|\mathbf{e}(l)\|^{\mathbf{w}}_{\text{block}}$
for Jacobi scheme and
$E^{\mathbf{w}}_{\text{block}}(t)=\frac{1-\alpha^K}{1-\alpha}\alpha^{t-1}\sum_{l=0}^{t-1}\alpha^{-l}\|\mathbf{e}(l)\|^{\mathbf{w}}_{\text{block}}$
for Gauss-Seidel scheme.

(b) If condition in (b) of Lemma \ref{Lem:contr-mapping-cov-error}
holds w.r.t. $\|\cdot\|^{\mathbf{w}}_{\text{block}}$, then
$E^{\mathbf{w}}_{\text{block}}(t)\leq \tilde
E^{\mathbf{w}}_{\text{block}} (t)$, where $\tilde
E^{\mathbf{w}}_{\text{block}} (t)\triangleq
\alpha^{t-1}\sum_{l=0}^{t-1}\alpha^{-l}\|\tilde{\mathbf{e}}_l\|^{\mathbf{w}}_{\text{block}}$
for Jacobi scheme and $\tilde E^{\mathbf{w}}_{\text{block}}
(t)\triangleq\frac{1-\alpha^K}{1-\alpha}
\alpha^{t-1}\sum_{l=0}^{t-1}\alpha^{-l}\|\tilde{\mathbf{e}}_l\|^{\mathbf{w}}_{\text{block}}$
for Gauss-Seidel scheme.

(c) If condition in (c) of Lemma \ref{Lem:contr-mapping-cov-error}
holds w.r.t. $\|\cdot\|^{\mathbf{w}}_{\text{block}}$, then
$E^{\mathbf{w}}_{\text{block}}(t)\leq
\bar{E}^{\mathbf{w}}_{\text{block}}(t)$, where
$\bar{E}^{\mathbf{w}}_{\text{block}}(t)\triangleq
\frac{1-\alpha^t}{1-\alpha}\|\bar{\mathbf{e}}\|^{\mathbf{w}}_{\text{block}}$
for Jacobi scheme with $\bar
E^{\mathbf{w}}_{\text{block}}(\infty)=\frac{1}{1-\alpha}\|\bar{\mathbf{e}}\|^{\mathbf{w}}_{\text{block}}$
and $\bar{E}^{\mathbf{w}}_{\text{block}}(t)\triangleq
\frac{1-\alpha^K}{1-\alpha}
\frac{1-\alpha^t}{1-\alpha}\|\bar{\mathbf{e}}\|^{\mathbf{w}}_{\text{block}}$
for Gauss-Seidel\footnote{Compared with Jacobi scheme, Gauss-Seidel
scheme and totally asynchronous scheme have extra error terms
$\frac{1-\alpha^K}{1-\alpha}$ and $\frac{1}{1-\alpha}$,
respectively, and
$\frac{1}{1-\alpha}>\frac{1-\alpha^K}{1-\alpha}>1$. } scheme with
$\bar
E^{\mathbf{w}}_{\text{block}}(\infty)=\frac{1-\alpha^K}{(1-\alpha)^2}\|\bar{\mathbf{e}}\|^{\mathbf{w}}_{\text{block}}$.
Furthermore, define the stationary set as $\mathbb
S^{\mathbf{w}}_{\text{block}} \triangleq \{\boldsymbol{\mathcal
Q}(\mathbf x)
:\|\mathbf{x}-\mathbf{x}^{*}\|^{\mathbf{w}}_{\text{block}}\leq \bar
E^{\mathbf{w}}_{\text{block}}(\infty)\}$. The sufficient condition
and necessary condition are the same as those in Lemma
\ref{Lem:contr-mapping-cov-error}.
\label{Lem:Jacobi-Gauss-Seidel-cov-error}
\end{Lem}
\begin{proof}
Please refer to Appendix B for the proof.
\end{proof}
\begin{Rem}
As a result of Lemma \ref{Lem:contr-mapping-cov-error} and Lemma
\ref{Lem:Jacobi-Gauss-Seidel-cov-error}, the effect of quantized
message passing affects the convergence property of the contracting
iterative algorithm in a fundamental way. From Lemma
\ref{Lem:Jacobi-Gauss-Seidel-cov-error}, the Jacobi and Gauss-Seidel
distributed iterative algorithms may not be able to converge
precisely to the fixed point under quantized message passing due to
the term $E^{\mathbf{w}}_{\text{block}}(t)$.
\end{Rem}

\section{Time Invariant Convergence-Optimal Quantizer Design}
\label{Sec:TI-COQ}

In this section,  we shall define a {\em Time Invariant Quantizer}
(TIQ) and then formulate the {\em Time Invariant Convergence-Optimal
Quantizer} (TICOQ) design problem. We shall consider the TICOQ
design for the scalar quantizer (SQ) and the vector quantizer (VQ)
cases separately. Specifically, the component quantizer $\mathcal
Q_k$ of the $k$-th node can be a group of scalar quantizers
$\mathcal Q_k^s=(\mathcal Q_m^s)_{m\in \mathcal M_k}$  or a vector
quantizer $\mathcal Q_k^v$. In the SQ case, each element
$T_m(\cdot)$ ($m \in \mathcal M_k$) of the vector $\mathbf
T_k(\cdot)$ is quantized by a coordinate scalar quantizer $Q_m^s$
separately. However, in the VQ case, the input to a vector quantizer
$Q_k^v$ is the vector $\mathbf T_k(\cdot)$.

\begin{Def}[Time Invariant Quantizer (TIQ)]
A {\em Time Invariant Quantizer} (TIQ) is a quantizer
$\mathbf{\mathcal{Q}}=(\mathbf{\mathcal{E}}, \mathbf{\mathcal{D}})$
such that $\mathbf{\mathcal{E}}$ and $\mathbf{\mathcal{D}}$ are time
invariant mappings. ~ \hfill\QED \label{def:TIQ}
\end{Def}

The system scalar TIQ  can be denoted as $\boldsymbol{\mathcal
Q}^s=(\mathcal Q_1^s,\cdots, \mathcal Q_m^s, \cdots, \mathcal
Q_n^s)$. Let $\mathbf{L}^s=(L_1^s,\cdots,L_n^s)$ be the {\em
quantization rate vector} for the system scalar TIQ
$\boldsymbol{\mathcal Q}^s$, where $L_m^s\in \mathbb Z^+$ is the
quantization rate (number of bits) of the coordinate scalar
quantizer $\mathcal Q_m^s$ ($1 \leq m \leq n$). The sum quantization
rate of the system scalar TIQ $\boldsymbol{\mathcal Q}^s$ is given
by $\sum_{m=1}^n L_m^s$.
Similarly, the system vector TIQ can be denoted as
$\boldsymbol{\mathcal Q}^v=(\mathcal Q_1^v,\cdots,\mathcal
Q_k^v,\cdots, \mathcal Q_K^v)$. Let
$\mathbf{L}^v=(L_1^v,\cdots,L_K^v)$ be the {\em quantization rate
vector} for the system vector TIQ $\boldsymbol{\mathcal Q}^v$, where
$L_k^v\in \mathbb Z^+$ is defined as the quantization rate (number
of bits) of the coordinate vector quantizer $\mathcal Q_k^v$ ($1\leq
k \leq K$). The sum quantization rate of the system vector TIQ
$\boldsymbol{\mathcal Q}^v$ is given by $\sum_{k=1}^K L_k^v$.

Using Lemma \ref{Lem:Jacobi-Gauss-Seidel-cov-error} (c), the
limiting error bound of the algorithm trajectory is given by
$\bar{E}^{\mathbf{w}}_{\text{block}}(\infty)=
\frac{1}{1-\alpha}\|\bar{\mathbf{e}}\|^{\mathbf{w}}_{\text{block}}$
(Jacobi scheme) or $\bar{E}^{\mathbf{w}}_{\text{block}}(\infty)
=\frac{1-\alpha^K}{(1-\alpha)^2}\|\bar{\mathbf{e}}\|^{\mathbf{w}}_{\text{block}}$
(Gauss-Seidel scheme). Therefore, the TICOQ design, which minimizes
$\bar{E}^{\mathbf{w}}_{\text{block}}(\infty)$ under the sum
quantization rate constraint, is equivalent to the following:
\begin{Prob}[TICOQ Design Problem]
\begin{align}
\min_{\boldsymbol{\mathcal Q}^s \text{or} \boldsymbol{\mathcal Q}^v
}   & \|\bar{\mathbf{e}}\|^{\mathbf{w}}_{\text{block}}
\label{eqn:TICOQ-obj}\\
s.t.  &  \sum_{m=1}^n L_m^s =L,\  L_m^s \in \mathbb Z^+
(1\leq m \leq n)  \ \text{SQ} \label{eqn:SQ-sum-rate-cons}  \\
\text{or}  & \sum_{k=1}^K L_k^v =L, \  L_k^v \in \mathbb Z^+ (1\leq
k \leq K)  \ \text{VQ} \label{eqn:VQ-sum-rate-cons}
\end{align}
where $\|\bar{\mathbf{e}}\|^{\mathbf{w}}_{\text{block}} =
\max_{\mathbf{x}\in \boldsymbol{\mathcal{X}}}
\|\mathbf{x}-\boldsymbol{\mathcal{Q}}^s(\mathbf{x})\|^{\mathbf{w}}_{\text{block}}$
(SQ case) or $\|\bar{\mathbf{e}}\|^{\mathbf{w}}_{\text{block}} =
\max_{\mathbf{x}\in \boldsymbol{\mathcal{X}}}
\|\mathbf{x}-\boldsymbol{\mathcal{Q}}^v(\mathbf{x})\|^{\mathbf{w}}_{\text{block}}$
(VQ case). \label{Prob:TICOQ}
\end{Prob}

\begin{Rem} [Interpretation of Problem \ref{Prob:TICOQ}] Note that the
optimization variable in Problem \ref{Prob:TICOQ} is the system TIQ
$\boldsymbol{\mathcal Q}^s$ or  $\boldsymbol{\mathcal Q}^v$. The
objective function $\|\bar{\mathbf{e}}\|^{\mathbf{w}}_{\text{block}}
= \max_{\mathbf{x}\in \boldsymbol{\mathcal{X}}}
\|\mathbf{x}-\boldsymbol{\mathcal{Q}}^s(\mathbf{x})\|^{\mathbf{w}}_{\text{block}}$
 or $\|\bar{\mathbf{e}}\|^{\mathbf{w}}_{\text{block}} =
\max_{\mathbf{x}\in \boldsymbol{\mathcal{X}}}
\|\mathbf{x}-\boldsymbol{\mathcal{Q}}^v(\mathbf{x})\|^{\mathbf{w}}_{\text{block}}$
obviously depends on the choice of the system TIQ
$\boldsymbol{\mathcal Q}^s$ or $\boldsymbol{\mathcal Q}^v$.
Furthermore, the constraint \eqref{eqn:SQ-sum-rate-cons} or
\eqref{eqn:VQ-sum-rate-cons} is the constraint on the quantization
rate $\mathbf L^s=(L^s_1,\cdots, L^s_n)$ or $\mathbf
L^v=(L^v_1,\cdots, L^v_K)$, which is also an effective constraint on
the optimization domains of $\boldsymbol{\mathcal Q}^s$ or
$\boldsymbol{\mathcal Q}^v$, respectively. It is because $L^s_m$ or
$L^v_k$ is a parameter (corresponding to the cardinality of the
index set, i.e. $|\mathcal I^s_m|=2^{L^s_m}$ or $|\mathcal
I^v_k|=2^{L^v_k}$) of the encoder and decoder of $\mathcal Q^s_m$ or
$\mathcal Q^v_k$. The Lagrangian function of Problem
\ref{Prob:TICOQ} is given by: $\mathcal L^s(\boldsymbol{\mathcal
Q}^s,\mu^s)=\|\bar{\mathbf{e}}\|^{\mathbf{w}}_{\text{block}}+\mu^s(\sum_{m=1}^n
L_m^s-L)$ (SQ case) or $\mathcal L^v(\boldsymbol{\mathcal
Q}^v,\mu^v)=\|\bar{\mathbf{e}}\|^{\mathbf{w}}_{\text{block}}+\mu^v(\sum_{k=1}^K
L_k^v -L)$ (VQ case), where $\mu^s$ or $\mu^v$ is the Lagrange
multiplier (LM) corresponding to the constraint
\eqref{eqn:SQ-sum-rate-cons} or \eqref{eqn:VQ-sum-rate-cons}. Hence,
the optimization problem \ref{Prob:TICOQ} can also be interpreted as
optimizing the tradeoff between the convergence performance
$\|\bar{\mathbf{e}}\|^{\mathbf{w}}_{\text{block}}$ and the
communication overhead $\sum_{m=1}^n L_m^s$ or $\sum_{k=1}^K L_k^v$.
The LM $\mu^s$ or $\mu^v$ can be regarded as the {\em per-iteration
cost sensitivity}.
\end{Rem}

\begin{Rem} [Robust Consideration in Problem \ref{Prob:TICOQ}]
The optimization objective
$\|\bar{\mathbf{e}}\|^{\mathbf{w}}_{\text{block}}$ in
\eqref{eqn:TICOQ-obj} actually corresponds to a {\em worst case
error} in the {\em algorthm trajectory}. In other words, the TICOQ
design is trying to find the optimal TIQ which minimizes the worst
case error. In fact, the algorithm trajectory $\mathbf{x}(t)$ is a
random process induced by the uncertainty in the initial point
$\mathbf{x}(0)$. In general, we do not have knowledge on the
distribution of $\mathbf{x}(t)$ due to the uncertainty on
$\mathbf{x}(0)$. Hence, the solution to Problem \ref{Prob:TICOQ}
(optimizing the worst case error) offers some robustness w.r.t. the
choice of $\mathbf{x}(0)$. ~ \hfill\QED
\end{Rem}

In the following, we shall discuss the scalar and vector TICOQ
design based on Problem \ref{Prob:TICOQ} separately.

\subsection{Time Invariant Convergence-Optimal Scalar Quantizer}
%
%

We first have a lemma on the structure of the optimizing quantizer
in the scalar TICOQ design in Problem \ref{Prob:TICOQ}.
\begin{Lem}[Structure of the Scalar TICOQ]
If each component norm $\|\cdot\|_k$ on $\mathbb{R}^{n_k}$ ($1\leq k
\leq K$) of the weighted block-maximum norm
$\|\cdot\|^{\mathbf{w}}_{\text{block}}$ defined by
\eqref{eqn:weighted-block-max-norm} is monotone (or absolute)
norm\footnote{A vector norm is monotone if and only if it is
absolute\cite{matrixanalysis:1986}.}, then the optimal coordinate
scalar quantizer $\mathcal Q_m^{s*}$ ($1 \leq m \leq n$) w.r.t. the
worst case error $\|\bar{\mathbf{e}}\|^{\mathbf{w}}_{\text{block}} =
\max_{\mathbf{x}\in \boldsymbol{\mathcal{X}}}
\|\mathbf{x}-\boldsymbol{\mathcal{Q}}^s(\mathbf{x})\|^{\mathbf{w}}_{\text{block}}$
is a uniform quantizer. \label{Lem:SQ}
\end{Lem}

\begin{proof}
Please refer to Appendix C for the proof.
\end{proof}


While the optimization variable in Problem \ref{Prob:TICOQ} (SQ
case) is $\boldsymbol{\mathcal Q}^s=(Q^s_1,\cdots,Q^s_n)$, using
Lemma \ref{Lem:SQ}, we can restrict the optimization domain of each
coordinate scalar quantizer $\mathcal Q^s_m$ ($1\leq m \leq n$) to
uniform quantizer without loss of optimality. Thus, the worst case
error of the $m$-th coordinate is given by $|\bar
e_m|\triangleq\max_{x_m \in \mathcal X_m}|x_m -\mathcal Q^s_m(x_m)
|=\frac{|\mathcal X_m|}{2\times 2^{L^s_m}}$ ($1\leq m \leq n$),
where $|\mathcal X_m|$ is the length of the interval $\mathcal X_m$
($x_m \in \mathcal X_m$), and the remaining optimization variable is
reduced from $\boldsymbol{\mathcal Q}^s=(Q^s_1,\cdots,Q^s_n)$ to
$\mathbf{L}^s=(L_1^s,\cdots,L_n^s)$. Scalar TICOQ design in Problem
\ref{Prob:TICOQ} w.r.t. $\mathbf{L}^s=(L_1^s,\cdots,L_n^s)$ is a
{\em Nonlinear Integer Programming} (NLIP) problem, which is in
general difficult to solve. Verifying the optimality of a solution
requires enumerating all the feasible solutions in most cases. In
the following, we shall derive the optimal solution to the scalar
TICOQ design in Problem \ref{Prob:TICOQ} w.r.t. the weighted
block-maximum norm $\|\cdot\|^{\mathbf{w}}_{\text{block}}$ defined
by \eqref{eqn:weighted-block-max-norm}, in which each component norm
$\|\cdot\|_k$ is the weighted maximum norm and the $L_p$ norm
separately.

\begin{Thm}[Solution for Weighted Maximum Norm]
Given a weighted block-maximum norm
$\|\cdot\|^{\mathbf{w}}_{\text{block}}$ defined by
\eqref{eqn:weighted-block-max-norm} (parameterized by
$\mathbf{w}=(w_1,\cdots,w_K)$) with $\|\cdot\|_k$ being the weighted
maximum norm $\|\cdot\|^{\mathbf a_k}_{\infty}$ defined by
\eqref{eqn:max-weight-norm-def}  (parameterized by $\mathbf
a_k=(a_m)_{m \in \mathcal M_k}$), let $\bar{L}_m^{s*}=(\log_2
\frac{C_m}{\tau})^{+}$, where $C_m\triangleq \frac{|\mathcal
X_m|}{2a_m(\sum_{k=1}^K w_k\mathbf{I}[m\in \mathcal M_k])}$,
$\mathbf{I}[\cdot]$ is indicator function and $\tau>0$ is a constant
related to the LM of the constraint \eqref{eqn:SQ-sum-rate-cons}
chosen to satisfy the constraint $\sum_{m=1}^n
(\log_2\frac{C_m}{\tau})^{+}=L$. The optimal integer solution of
Problem \ref{Prob:TICOQ} for the SQ case  is given by\footnote{We
arrange the real sequence $z_1,\cdots,z_n$ in decreasing order and
denote them as $z_{[1]}\geq\cdots \geq z_{[m]}\geq\cdots\geq
z_{[n]}$, where $z_{[m]}$ represents the $m$-th largest term of
$\{z_m\}$.}:
\begin{align}
L_{[m]}^{s*}=\left\{
\begin{array}{ll}
\lceil \bar{L}_{[m]}^{s*} \rceil, &
\text{if $m\leq  \sum_{m=1}^n(\bar{L}_m^{s*}-\lfloor \bar{L}_m^{s*} \rfloor)$} \\
\lfloor \bar{L}_{[m]}^{s*} \rfloor, & \text{otherwise}
\end{array} \right. \label{eqn:sol-SQ}
\end{align}
The optimal value of Problem \ref{Prob:TICOQ} under continuous
relaxation is $\tau$.
 \label{Lem:sol-SQ-weighted}
\end{Thm}
\begin{proof}
Please refer to Appendix D for the Proof.
\end{proof}

\begin{Thm}[Solution for $L_p$ Norm] Given a weighted block-maximum norm
$\|\cdot\|^{\mathbf{w}}_{\text{block}}$ defined by
\eqref{eqn:weighted-block-max-norm} (parameterized by
$\mathbf{w}=(w_1,\cdots,w_K)$) with $\|\cdot\|_k$ being the $L_p$
norm $\|\cdot\|_{p}$ defined by \eqref{eqn:L_p-norm-def}
(parameterized by $p$), the optimal solution of Problem
\ref{Prob:TICOQ} for the SQ case with continuous relaxation ($L_m^s
\in \mathbb R^+$) is $\bar L_m^{s*}= \frac{1}{p}\log_2 (\frac{
C_m}{\sum_{k=1}^K \tau_k \mathbf I[m \in \mathcal M_k]}\vee 1)$,
where $C_m\triangleq \frac{|\mathcal X_m|^p}{2^p (\sum_{k=1}^K
w_k^p\mathbf{I}[m\in M_k])}$, $\{\tau_1,\cdots,\tau_K\}$ and $\tau$
are constants related to the LM of the constraint
\eqref{eqn:VQ-sum-rate-cons} chosen to satisfy the constraint
$\sum_{k=1}^K \sum_{m\in \mathcal M_k} \frac{1}{p}\log_2 (\frac{
C_m}{ \tau_k }\vee 1)=L$ and complementary slackness conditions
$\frac{1}{\tau_k}\big(\sum_{m\in \mathcal M_k}(C_m \wedge
\tau_k)-\tau \big)=0$ ($\forall k\in \{1,\cdots,
K\}$)\footnote{$x\vee a \triangleq \max\{x,a\}$ and $x\wedge a
\triangleq \min\{x,a\}$.}. The optimal value of Problem
\ref{Prob:TICOQ} with continuous relaxation is $\tau^{\frac{1}{p}}$.
\label{Lem:sol-SQ-Lp}
\end{Thm}
\begin{proof}
Please refer to Appendix D for the Proof.
\end{proof}

\begin{Rem} [Determination of $\{\tau_1,\cdots, \tau_K\}$ and
$\tau$] Solving for $\{\tau_1,\cdots, \tau_K\}$ and $\tau$ involves
solving a system of $K+1$ equations with $K+1$ unknowns. We have
$2^K-1$ valid cases for the above system of equations according to
$\tau_k>\max_{m\in \mathcal M_k} C_m$ or $\tau_k \leq \max_{m\in
\mathcal M_k}C_m$ ($\forall k$). Firstly, if $\tau_k>\max_{m\in
\mathcal M_k}C_m$ ($\forall k$), then $\bar L^s_m=0$ ($\forall m$),
which is not a valid case. Therefore, without loss of generality,
assume $\tau_k\leq \max_{m\in \mathcal M_k}C_m$ $\forall k\in
\{1,\cdots,N\}$ and $\tau_k>\max_{m\in \mathcal M_k}C_m$ $\forall
k\in \{N+1,\cdots,K\}$. The system of $K+1$ equations and unknowns
reduce to $N+1$ equations, which are given by: $\sum_{k=1}^N
\sum_{m\in \mathcal M_k}\frac{1}{p}\log_2(\frac{C_m}{\tau_k}\vee
1)=L$ and $\sum_{m \in \mathcal M_k}(C_m \wedge \tau_k)=\tau$
($\forall k \in \{1,\cdots, N\}$), and $N+1$ unknowns\footnote{For
example, consider $K=2$, $n=4$, $\mathcal M_1=\{1,2\}$ and $\mathcal
M_2=\{3,4\}$ . We have $2^2-1=3$ valid cases. Case 1 ( $\tau_1 \leq
\max_{m\in \mathcal M_1}C_m$ and $\tau_2 \leq \max_{m\in \mathcal
M_2}C_m$ ) We have 3 equations: $\sum_{k=1}^2 \sum_{m\in \mathcal
M_k}\frac{1}{p}\log_2(\frac{C_m}{\tau_k}\vee 1)=L$, $\sum_{m \in
\mathcal M_k}(C_m \wedge \tau_k)=\tau$ ($\forall k \in \{1,2\}$),
and 3 unknowns: $\{\tau_1, \tau_2\}$,$\tau$. Case 2 ( $\tau_1 \leq
\max_{m\in \mathcal M_1}C_m$ and $\tau_2 > \max_{m\in \mathcal
M_2}C_m$ ) We have 2 equations: $ \sum_{m\in \mathcal
M_1}\frac{1}{p}\log_2(\frac{C_m}{\tau_1}\vee 1)+0+0=L$,$\sum_{m \in
\mathcal M_1}(C_m \wedge \tau_1)=\tau$, and 2 unknowns: $\tau_1$,
$\tau$. Case 3 ( $\tau_1 >\max_{m\in \mathcal M_1}C_m$ and
$\tau_2\leq \max_{m\in \mathcal M_2}C_m$ ) We have 2 equations:
$0+0+ \sum_{m\in \mathcal
M_2}\frac{1}{p}\log_2(\frac{C_m}{\tau_2}\vee 1)=L$,$\sum_{m \in
\mathcal M_2}(C_m \wedge \tau_2)=\tau$, and 2 unknowns: $\tau_2$,
$\tau$.}: $\{\tau_1,\cdots,\tau_N\}$ and $\tau$.
\end{Rem}

\subsection{Time Invariant Convergence-Optimal Vector Quantizer}

%
%

We first have a lemma on the structure of the optimizing quantizer
in the vector TICOQ design in Problem \ref{Prob:TICOQ}.

\begin{Lem}[Structure of the Vector TICOQ]
If the component norm $\|\cdot\|_k$ $(1\leq k \leq K)$ on
$\mathbb{R}^{n_k}$ of the weighted block-maximum norm
$\|\cdot\|^{\mathbf{w}}_{\text{block}}$ defined by
\eqref{eqn:weighted-block-max-norm} is monotone (or absolute) norm,
each  vector TICOQ $\mathcal Q_k^{v*}$ is a $n_k$-dimensional
lattice quantizer, the structure of which is uniquely determined by
the norm $\|\cdot\|_k$ on $\mathbb{R}^{n_k}$. In particular, if
$\|\cdot\|_k$ ($1\leq k \leq K$) is $L_2$ norm, each  vector TICOQ
$\mathcal Q_k^{v*}$ is the thinnest lattice for the covering problem
in Euclidean space; If $\|\cdot\|_k$ ($1\leq k \leq K$) is weighted
maximum norm, each vector TICOQ $\mathcal Q_k^{v*}$ reduces to $n_k$
coordinate scalar TICOQ $\mathcal Q^{s*}_m$ ($m\in \mathcal M_k$)
with scalar quantization\footnote{In other words, the vector TICOQ
design reduces to scalar TICOQ design discussed in Theorem
\ref{Lem:sol-SQ-weighted}.} of each coordinate $x_m$ ($m\in \mathcal
M_k$) of $\mathbf{x}_k$. \label{Lem:VQ}
\end{Lem}
\begin{proof}
Please refer to Appendix E for the proof.
\end{proof}

A lattice is a regular arrangement of points in $n$-dimensional
space that includes the origin. ``Regular'' means that each point
``sees'' the same geometrical environment as any other point
\cite{GrayQuantizationBook:1992}. The lattice covering problem asks
for the most economical way to arrange the lattice points so that
the $n$-dimensional space can be covered with overlapping spheres
whose centers are the lattice points, i.e. tries to find the
thinnest (i.e. minimum density\footnote{The density of a covering is
the defined as the number of spheres that contain a point of the
space\cite{ConvwaySloaneLatticebook:1999}.}) lattice
covering\cite{ConvwaySloaneLatticebook:1999}.  The thinnest lattice
coverings known in all dimension $n$ ($n\leq 23$) are the dual
lattice $A^*_n$ (when $1 \leq n \leq 5$, $A^*_n$ is known to be
optimal)\cite{ConvwaySloaneLatticebook:1999}. The worst case error
of the dual lattice $A^*_{n_k}$ for the $k$-th node when
$\|\cdot\|_k$ is $L_p$ norm ($p>2$) is less than that measured in
$L_2$ norm (Appendix E). Therefore, if $\|\cdot\|_k$ is $L_p$ norm
($p\geq2$), we can solve the TICOQ design (VQ case) in Problem
\ref{Prob:TICOQ} using dual lattice structure $A^*_{n_k}$. Thus, the
worst case error is given by
$\|\bar{\mathbf{e}}_k\|_k=\Big(\frac{\prod_{m \in \mathcal
M_k}|\mathcal
X_m|}{\sqrt{\frac{1}{n_k+1}}}\Big)^{\frac{1}{n_k}}\sqrt{\frac{n_k(n_k+2)}{12(n_k+1)}}2^{-\frac{L^v_k}{n_k}}$
(Appendix E), where $|\mathcal X_m|$ is the length of the interval
$\mathcal X_m$ ($x_m \in \mathcal X_m$) ($1\leq m \leq n$), and the
remaining optimization variable is reduced from
$\boldsymbol{\mathcal Q}^v=(Q^v_1,\cdots,Q^v_K)$ to
$\mathbf{L}^v=(L_1^v,\cdots,L_K^v)$. Similarly, vector TICOQ design
in Problem \ref{Prob:TICOQ} w.r.t.
$\mathbf{L}^v=(L_1^v,\cdots,L_K^v)$ is also a {\em Nonlinear Integer
Programming} (NLIP) problem, which is in general difficult to solve.
In the following, we shall derive the optimal solution to the vector
TICOQ design in Problem \ref{Prob:TICOQ} based on dual lattice
$A^*_{n_k}$.

\begin{Thm}[Solution for Dual Lattice quantizer] Given a weighted block-maximum norm
$\|\cdot\|^{\mathbf{w}}_{\text{block}}$ defined by
\eqref{eqn:weighted-block-max-norm} (parameterized by
$\mathbf{w}=(w_1,\cdots,w_K)$) and the dual lattice $\{A^*_{n_k}:
1\leq k \leq K\}$ quantizer, let $\bar{L}_k^{v*}=n_k(\log_2
\frac{D_k}{\tau})^{+}$, where $D_k\triangleq
\frac{1}{w_k}\Big(\frac{\prod_{m \in \mathcal M_k}|\mathcal
X_m|}{\sqrt{\frac{1}{n_k+1}}}\Big)^{\frac{1}{n_k}}\sqrt{\frac{n_k(n_k+2)}{12(n_k+1)}}$,
and $\tau>0$ is a constant related to the LM of the constraint
\eqref{eqn:VQ-sum-rate-cons} chosen to satisfy the constraint
$\sum_{k=1}^K n_k(\log_2 \frac{D_k}{\tau})^{+}=L$. The optimal
integer solution of Problem \ref{Prob:TICOQ} for the VQ case w.r.t.
dual lattice $\{A^*_{n_k}: 1\leq k \leq K\}$ quantizer  when
$n_1=\cdots=n_K$ is given by:
\begin{align}
L_{[k]}^{v*}=\left\{
\begin{array}{ll}
\lceil \bar{L}_{[k]}^{v*} \rceil, &
\text{if $k\leq \sum_{k=1}^K (\bar{L}_k^{v*}-\lfloor \bar{L}_k^{v*} \rfloor)$} \\
\lfloor \bar{L}_{[k]}^{v*} \rfloor, & \text{otherwise}
\end{array} \right. \label{eqn:sol-VQ}
\end{align}
The optimal value of Problem \ref{Prob:TICOQ} under continuous
relaxation is $\tau$. \label{Lem:sol-VQ-L2}
\end{Thm}
\begin{proof}
Please refer to Appendix E for the Proof.
\end{proof}

\subsection{Tradeoff between Convergence Error and Message Passing Overhead}
In this subsection, we shall quantify the tradeoff between the
convergence error of the algorithm trajectory and the message
passing overhead using TICOQ. Specifically, the steady-state
convergence error in the algorithm trajectory is related to
$\bar{E}^{\mathbf{w}}_{\text{block}}(\infty)$ and the message
passing overhead is related to the sum quantization rate (number of
bits) $L$ of the system quantizer. The following lemma summarizes
the tradeoff result.


\begin{Lem}(\emph{Performance Tradeoff of the Scalar and Vector TICOQ}) For $L\geq
L'$ ($L\in \mathbb Z^{+}$), where
\begin{align}
L'=\left\{
\begin{array}{ll}
\sum_{m=1}^n \log_2 C_m-n \log_2 (\min_m C_m),
\text{WM norm} \\
\sum_{m=1}^n \log_2 \tilde C_m -n \log_2 \big( \min_m \tilde C_m
\big),   \text{$L_p$ norm}\\
\sum_{k=1}^K \log_2 D_k- K\log_2 (\min_k D_k),  \text{dual lattice}
\end{array} \right. \label{eqn:sol-VQ}
\end{align}
$\tilde C_m= ( \sum_{k=1}^K n_k \mathbf 1 [m \in \mathcal M_k])
C_m$, the limiting error bound of the scalar and vector TICOQ
considered in this section is given by
$\bar{E}^{\mathbf{w}}_{\text{block}}(\infty)
=\frac{1}{1-\alpha}\mathcal{O}(2^{-\frac{L}{n}})$.\label{Lem:performace-tradeoff-invariant}
\end{Lem}
\begin{proof}
Please refer to Appendix F for the proof.
\end{proof}
\begin{Rem}
As $L\to \infty$, $\bar{E}^{\mathbf{w}}_{\text{block}}(\infty)\to
0$, which reduces to the conventional convergence results with
perfect message passing. On the other hand,
$\bar{E}^{\mathbf{w}}_{\text{block}}(\infty)>0$ for finite $L$ and
hence, we cannot guarantee the convergence behavior of the
contracting iterations with TICOQ to the fixed point $\mathbf x^*$
of the contraction mapping $\mathbf T$ anymore. Nevertheless, the
convergence error decreases exponentially w.r.t. the message passing
overhead $L$. ~ \hfill\QED
\end{Rem}

\section{Time Varying Convergence-Optimal Quantizer Design}
\label{Sec:TV-COQ}

Similar to Section \ref{Sec:TV-COQ},  we shall define a {\em Time
Varying Quantizer} (TVQ)  and then formulate the {\em Time Varying
Convergence-Optimal Quantizer} (TVCOQ) design problem. We shall
consider the TVCOQ design for both the SQ and VQ cases separately.

\begin{Def}[Time Varying Quantizer (TVQ)]
A {\em Time Varying Quantizer} (TVQ) is a quantizer
$\mathbf{\mathcal{Q}}(t)=\big(\mathbf{\mathcal{E}}(t),
\mathbf{\mathcal{D}}(t)\big)$ such that $\mathbf{\mathcal{E}}(t)$
and $\mathbf{\mathcal{D}}(t)$ changes with time. In other words, the
quantizer $\mathbf{\mathcal{Q}}(t)=\big(\mathbf{\mathcal{E}}(t),
\mathbf{\mathcal{D}}(t)\big)$ at the $t$-th iteration is function of
time $t$. ~ \hfill\QED \label{def:TVQ}
\end{Def}

The system  scalar TVQ at the $t$-th iteration can be denoted as
$\boldsymbol{\mathcal Q}^s(t)=\big(\mathcal Q_1^s(t),\cdots,
\mathcal Q_m^s(t), \cdots, \mathcal Q_n^s(t)\big)$ with quantization
rate vector (at the $t$-th iteration)
$\mathbf{L}^s(t)=\big(L_1^s(t),\cdots,L_n^s(t)\big)$. Similarly, the
system vector TVQ at the $t$-th iteration can be denoted as
$\boldsymbol{\mathcal Q}^v(t)=\big(\mathcal Q_1^v(t),\cdots,\mathcal
Q_k^v(t),\cdots, \mathcal Q_K^v(t)\big)$  with quantization rate
vector (at the $t$-th iteration)
$\mathbf{L}^v(t)=\big(L_1^v(t),\cdots,L_K^v(t)\big)$.
Using the result (c) of Lemma
\ref{Lem:Jacobi-Gauss-Seidel-cov-error}, the TVCOQ design, which
minimize $\tilde{E}^{\mathbf{w}}_{\text{block}}(\bar T)$ under total
quantization rate constraint over a horizon of $\bar T$ iterations,
is equivalent to the following:

\begin{Prob}[TVCOQ Design Problem]
\begin{align}
& \min_{\substack{\text{ }\{\boldsymbol{\mathcal Q}^s(t):0\leq t
\leq \bar T\}\\\text{or}\{\boldsymbol{\mathcal Q}^v(t):0\leq t \leq
\bar T\}}}    \alpha^{\bar T-1}\sum_{t=0}^{\bar
T-1}\alpha^{-t}\|\tilde{\mathbf{e}}_t\|^{\mathbf{w}}_{\text{block}}
\label{eqn:TVCOQ-obj}\\
& \ s.t.   \sum_{t=0}^{\bar T-1}\sum_{m=1}^n L_m^s(t) =\bar T L,
L_m^s(t) \in \mathbb Z^+  (\forall m, t)
\text{SQ} \label{eqn:TV-VQ-sum-rate-cons}\\
& \ \ \text{or}  \sum_{t=0}^{\bar T-1}\sum_{k=1}^K L_k^v(t) =\bar T
L,
 L_k^v(t) \in \mathbb Z^+ (\forall k,t) \text{VQ} \label{eqn:TV-VQ-sum-rate-cons}
\end{align}  \label{Prob:TVCOQ}
\end{Prob}


By introducing additional auxiliary variables $\{L(t): 0 \leq t \leq
\bar T-1\}$, where $L(t)$ can be interpreted as the per-stage sum
quantization rate, and applying primal decomposition techniques, we
can decompose Problem \ref{Prob:TVCOQ} into subproblems (per-stage
TICOQ design $\boldsymbol{\mathcal Q}^s(t)$ or $\boldsymbol{\mathcal
Q}^v(t)$ ($0\leq t \leq \bar T-1$)), which are given by
\begin{Prob}(TVCOQ Subproblems: Per-stage TICOQ Design Problem)
\begin{align}
& \min_{\boldsymbol{\mathcal Q}^s(t)\text{or}\boldsymbol{\mathcal
Q}^v(t)}
\|\tilde{\mathbf{e}}_t\big(L(t)\big)\|^{\mathbf{w}}_{\text{block}}
\label{eqn:TVCOQ-subp-obj}\\
& s.t. \ \   \sum_{m=1}^n L_m^s(t) =L(t),  L_m^s(t) \in \mathbb Z^+
(\forall m) \text{SQ}
\label{eqn:TV-per-stage-SQ-sum-rate-cons}\\
& \text{or} \ \   \sum_{k=1}^K L_k^v(t) =L(t),  L_k^v(t) \in \mathbb
Z^+ (\forall k ) \text{VQ}\label{eqn:TV-per-stage-VQ-sum-rate-cons}
\end{align}
\label{Prob:TVCOQ-subp}
\end{Prob}
and the master problem (per-stage sum quantization rate $\{L(t): 0
\leq t \leq \bar T-1\}$ allocation among stages), which is given by
\begin{Prob}(TVCOQ Master Problem:  Sum Quantization Rate Allocation
Problem)
\begin{align}
& \min_{\{L(t): 0 \leq t \leq \bar T-1\}} \   \alpha^{\bar
T-1}\sum_{t=0}^{\bar
T-1}\alpha^{-t}\|\tilde{\mathbf{e}}_t^*\big(L(t)\big)\|^{\mathbf{w}}_{\text{block}}
\label{eqn:TVCOQ-masterp-obj}\\
& s.t. \  \sum_{t=0}^{\bar T-1} L(t)=\bar T L,\ L(t) \in \mathbb Z^+
(0 \leq t \leq \bar T-1) \label{eqn:TV-per-stage-sum-rate-allo}
\end{align}
\label{Prob:TVCOQ-masterp}
\end{Prob}
where
$\|\tilde{\mathbf{e}}_t^*\big(L(t)\big)\|^{\mathbf{w}}_{\text{block}}$
is the optimal  value of the $t$-th subproblem in Problem
\ref{Prob:TVCOQ-subp} for given $L(t)$ ($0 \leq t \leq \bar T-1$).
Given the per-stage sum quantization rate $L(t)$, each {\em
Per-stage TICOQ Design Problem} in Problem \ref{Prob:TVCOQ-subp} is
the same as the TICOQ design problem in Section \ref{Sec:TI-COQ},
and hence, the approaches in Section \ref{Sec:TI-COQ} can be applied
to solve Problem \ref{Prob:TVCOQ-subp} for given $L(t)$ ($0 \leq t
\leq \bar T-1$), including both SQ case and VQ case. In the
following, we shall mainly discuss the {\em Sum Quantization Rate
$\{L(t): 0 \leq t \leq \bar T-1\}$ Allocation} in Problem
\ref{Prob:TVCOQ-masterp} and analyze the tradeoff between
convergence error and message passing overhead for the TVCOQ design.

\subsection{Time Varying Convergence-Optimal Scalar and Vector Quantizers}

Similar to the TICOQ in Problem \ref{Prob:TICOQ}, each TVCOQ
subproblem in Problem \ref{Prob:TVCOQ-subp} is  a {\em Nonlinear
Integer Programming} (NLIP) problem. Brute-force solution to the
TVCOQ master problem in Problem \ref{Prob:TVCOQ-masterp} requires
exhaustive search, which is not acceptable. Therefore, we first
apply continuous relaxation to the subproblems in Problem
\ref{Prob:TVCOQ-subp} to obtain the closed-form expression
$\|\tilde{\mathbf{e}}_t^*\|^{\mathbf{w}}_{\text{block}}$ of the
$t$-th subproblem.  Based on the closed-form
$\|\tilde{\mathbf{e}}_t^*\|^{\mathbf{w}}_{\text{block}}$, the master
problem in Problem \ref{Prob:TVCOQ-masterp} becomes tractable, the
solution of which is summarized in the following lemma.

\begin{Thm}(Solution to TVCOQ Master Problem for SQ and
VQ) For any given $\bar T$, assume $L\geq L'-n \frac{\bar T-1}{2}
\log_2 \alpha $ ($L \in \mathbb Z^+$). Let $\bar L^*(t)=n
\log_2(\frac{\alpha^{-t}\ln 2 }{n \mu})$, where $\mu>0$ is the LM of
the constraint \eqref{eqn:TV-per-stage-sum-rate-allo} chosen to
satisfy the constraint $\sum_{t=0}^{\bar T-1} n
\log_2(\frac{\alpha^{-t}\ln 2 }{n \mu})=\bar T L$. The optimal
integer solution to the TVCOQ Master Problem in Problem
\ref{Prob:TVCOQ-masterp}  is given by
\begin{align}
L^*([t])=\left\{
\begin{array}{ll}
\lceil \bar{L}^*([t]) \rceil, &
\text{if $t\leq \sum_{t=0}^{\bar T-1} (\bar{L}^*(t)-\lfloor \bar{L}^*(t) \rfloor)$} \\
\lfloor \bar{L}^*([t]) \rfloor, & \text{otherwise}
\end{array} \right. \label{eqn:sol-masterp-int}
\end{align}
\label{Lem:sol-TVCOQ-masterp}
\end{Thm}
\begin{proof}
Please refer to Appendix G for the Proof.
\end{proof}

Given the per-stage sum quantization rate $\{L^*(t): 0 \leq t \leq
\bar T-1\}$ allocation obtained by Theorem
\ref{Lem:sol-TVCOQ-masterp}, the TVCOQ Subproblems in Problem
\ref{Prob:TVCOQ-subp} (similar to the TICOQ design problem in
Section \ref{Sec:TI-COQ}) can be easily solved by Theorem
\ref{Lem:sol-SQ-weighted}, \ref{Lem:sol-SQ-Lp} and
\ref{Lem:sol-VQ-L2}.

\subsection{Tradeoff between Convergence Error and Message Passing Overhead}

In this subsection, we shall quantify the tradeoff between the error
of the algorithm trajectory and the message passing overhead using
TVCOQ. The following lemma summarizes the tradeoff results.

\begin{Lem}(Performance Tradeoff of the Scalar and Vector TVCOQ) For
any given $\bar T$,  the convergence error bound (at the $\bar T$-th
iteration) of the scalar and vector TVCOQ  is given by 
$\tilde{E}^{\mathbf{w}}_{\text{block}}(\bar T)=\bar
T\alpha^{\frac{\bar T-1}{2}}\mathcal{O}(2^{-\frac{L}{n}})$ for
$L\geq L'-n \frac{\bar T-1}{2} \log_2 \alpha$ ($L \in \mathbb Z^+$),
where $L'$ is given by
\eqref{eqn:sol-VQ}.\label{Lem:performace-tradeoff-variant}
\end{Lem}
\begin{proof}
Please refer to Appendix G for the proof.
\end{proof}
\begin{Rem}
As $L\to \infty$, $\tilde{E}^{\mathbf{w}}_{\text{block}}(\bar T)\to
0$, which reduces to the conventional convergence results with
perfect message passing. On the other hand, for any fixed $L$, as
$\bar T\rightarrow \infty$, we have
$\tilde{E}^{\mathbf{w}}_{\text{block}}(\bar T)\rightarrow 0$. Hence,
using TVCOQ, one could achieve asymptotically zero convergence error
even with finite $L$. ~ \hfill\QED
\end{Rem}

\section{Simulation Results and Discussions}
\label{Sec:Simulation}

In this section, we shall evaluate the performance of the proposed
{\em time invariant convergence-optimal quantizer} (TICOQ) and {\em
time varying convergence-optimal quantizer} (TVCOQ) for the
contracting iterations by simulations. We consider distributed
precoding updates in MIMO interference game with $K$ transmitter and
receiver pairs where the transmit convariance matrix $\mathbf P_k$
of the $k$-th transmitter is iteratively updated (and quantized)
according to \eqref{eqn:mimo-centrl-iter-error}. In the simulation,
we consider $K = 2$, 4, 8 noncooperative transmitter-receiver pairs
with $N=2,4$ transmit/receive antenna. The distance from the $k$-th
transmitter to the $j$-th receiver is denoted as $d_{kj}$. The
bandwidth is 10 MHZ. The pathloss exponent is $\gamma=3.5$. Each
element of the small scale fading channel matrix is
$\mathcal{CN}(0,1)$ distributed. We compare the performance with two
reference baselines. Baseline 1 (BL1) refers to the case with
perfect message passing\cite{PalomarmuMIMOunifiedview:2008}.
Baseline 2 (BL2) refers to the case with uniform scalar quantizer.


%

\subsection{Performance of the TICOQ}

\begin{figure}[t]
\begin{center}
  \subfigure[Sum throughput]
  {\resizebox{7.5cm}{!}{\includegraphics{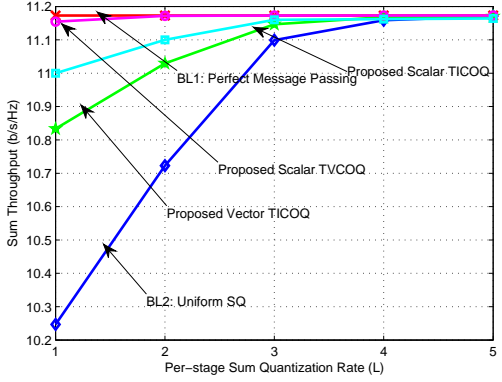}}}
  \subfigure[Convergence error]
  {\resizebox{7.5cm}{!}{\includegraphics{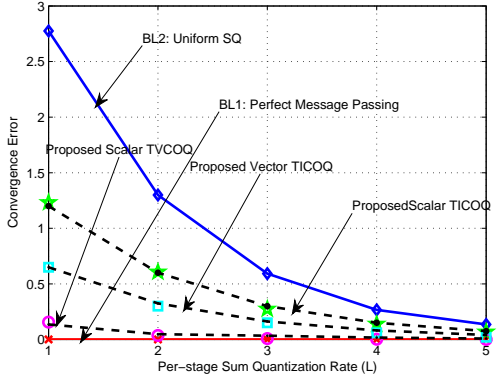}}}
  \end{center}
    \caption{Sum throughput/convergence error versus per-stage sum quantization rate $L$ (bits) of 2
pairs MIMO interference game with 2 transmit and receive antennas,
$d_{11}=d_{22}=100$ m, $d_{12}=200$ m, $d_{21}=500$ m, path loss
exponent $\gamma=3.5$, and transmit power of $P_1=P_2=10$ dBm. The
total number of iterations is $\bar T =4$ and the per-stage
quantization rate per antenna is $\frac{L}{K\times
N^2}=\frac{L}{8}$.  In (b), the ``*'', ``o'', etc represent the
simulation results of the proposed TICOQ and TVCOQ, while the dashed
line represents the analytical expression
$\mathcal{O}(2^{-\frac{L}{n}})$ (TICOQ) and $\bar
T\alpha^{\frac{\bar T-1}{2}}\mathcal{O}(2^{-\frac{L}{n}})$ at fixed
$\bar T=4$ (TVCOQ).}
    \label{Fig:vs_L}
\end{figure}

Fig. \ref{Fig:vs_L} (a) and Fig. \ref{Fig:vs_L} (b) illustrate the
sum throughput and convergence error (w.r.t. the weighted
block-maximum norm $\|\cdot\|^{\mathbf{w}}_{\text{block}}$) versus
the per-stage sum quantization rate of 2 pairs MIMO interference
game under fixed number of iterations. The sum throughput of all the
schemes increases as $L$ decreases due to the decreasing convergence
error. It can be observed that the proposed scalar and vector TICOQ
have significant performance gain in sum throughput and convergence
error compared with the commonly used uniform scalar quantizer. In
addition, the vector TICOQ outperforms the scalar TICOQ in
convergence performance at the cost of the higher encoding and
decoding complexity.

\begin{figure}[t]
\begin{center}
  \subfigure[Sum throughput]
  {\resizebox{7.5cm}{!}{\includegraphics{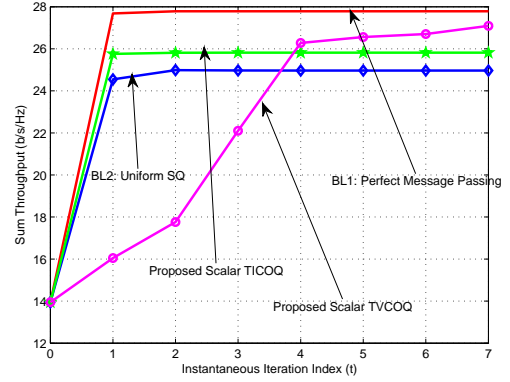}}}
  \subfigure[Convergence error]
  {\resizebox{7.5cm}{!}{\includegraphics{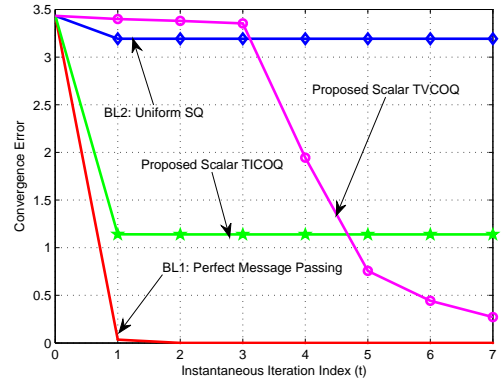}}}
  \end{center}
    \caption{Sum throughput/convergence error versus instantaneous iteration index $t$ of
    4
pairs MIMO interference game with 4 transmit and receive antennas,
$d_{ii}=100$ m, $d_{ij}=200$ m ($i<j$), $d_{ij}=500$ m ($i>j$), path
loss exponent $\gamma=3.5$, and transmit power of $P_1=P_2=5$ dBm.
The per-stage sum quantization rate $L=64$ bits (i.e. the per-stage
quantization rate per antenna is $\frac{L}{K\times
N^2}=\frac{L}{64}=1$ bit), and the total number of iterations $\bar
T=7$.}
    \label{Fig:4I_4N}
\end{figure}

\begin{figure}[t]
\begin{center}
  \subfigure[Sum throughput]
  {\resizebox{7.5cm}{!}{\includegraphics{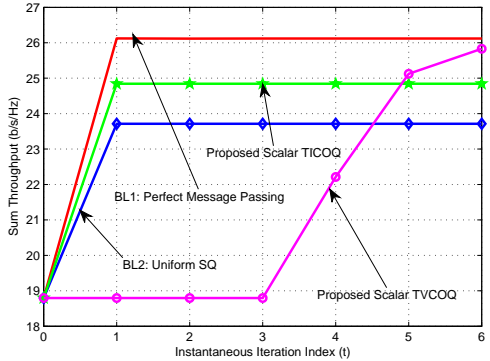}}}
  \subfigure[Convergence error]
  {\resizebox{7.5cm}{!}{\includegraphics{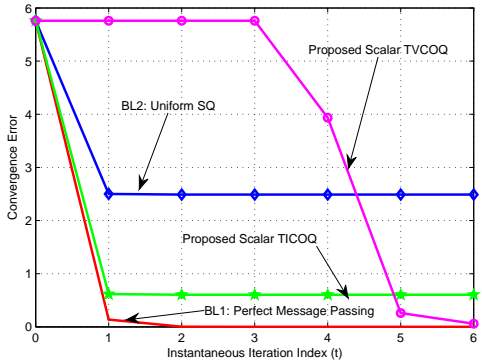}}}
  \end{center}
    \caption{Sum throughput/convergence error versus instantaneous iteration index $t$ of
    8
pairs MIMO interference game with 2 transmit and receive antennas,
$d_{ii}=100$ m, $d_{ij}=200$ m ($i<j$), $d_{ij}=500$ m ($i>j$), path
loss exponent $\gamma=3.5$, and transmit power of $P_1=P_2=5$ dBm.
The per-stage sum quantization rate $L=32$ bits (i.e. the per-stage
quantization rate per antenna is $\frac{L}{K\times
N^2}=\frac{L}{32}=1$ bit), and the total number of iterations $\bar
T=6$.}
    \label{Fig:8I}
\end{figure}

Fig. \ref{Fig:4I_4N} and Fig. \ref{Fig:8I} show the sum throughput
and convergence error versus the instantaneous iteration time index
of MIMO interference game  under a fixed per-stage sum quantization
rate and total number of iterations with different $K$ and $N$. It
can be seen that in all cases, the proposed scalar and vector TICOQ
have significant performance gain in sum throughput and convergence
error compared with the commonly used uniform scalar qunatizer.

\subsection{Performance of the TVCOQ}

From Fig. \ref{Fig:vs_L} (a) and Fig. \ref{Fig:vs_L} (b), we can
observe that the proposed TVCOQ has significant performance gain in
sum throughput and convergence error compared with commonly used
uniform scalar qunatizer and TICOQ. For example, the sum throughput
of the TVCOQ is very close to that with perfect message passing.
Fig. \ref{Fig:4I_4N} and \ref{Fig:8I} illustrate the transient
performance of the TVCOQ versus iteration index $t$. We observe that
the performance of the TVCOQ improves as $t$ increases. This is
because the TVCOQ optimizes the quantization rate allocation over
both the node domain and the time domain (over a horizon of $\bar T$
iterations).

\subsection{Tradeoff between Convergence Error and Message Passing Overhead}

Fig. \ref{Fig:vs_L} (b) illustrates the tradeoff between convergence
error and message passing overhead (in terms of  $L$ at fixed total
number of iterations $\bar T$). As the message passing overhead $L$
increases, the convergence error of all the proposed quantization
schemes approaches to 0 with order $\mathcal O (2^{-\frac{L}{n}})$
under fixed $\bar T$, which verified the results in Lemma
\ref{Lem:performace-tradeoff-invariant} and
\ref{Lem:performace-tradeoff-variant}. Similarly, the sum throughput
of all the schemes increases as $L$ decreases due to the decreasing
convergence error as shown in Fig. \ref{Fig:vs_L} (a).

\begin{figure}[t]
\begin{center}
  \subfigure[Sum throughput]
  {\resizebox{7.5cm}{!}{\includegraphics{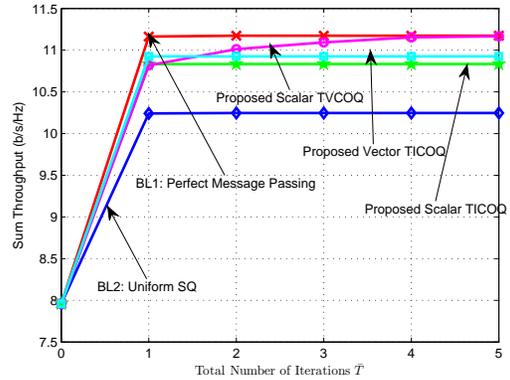}}}
  \subfigure[Convergence error]
  {\resizebox{7.5cm}{!}{\includegraphics{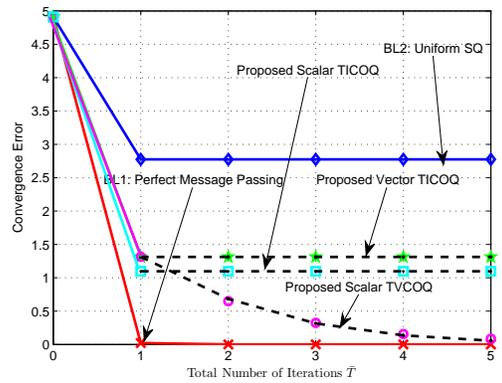}}}
  \end{center}
    \caption{Sum throughput/convergence error versus the total number of iterations $\bar T$ of 2
pairs MIMO interference game with 2 transmit and receive antennas,
$d_{11}=d_{22}=100$ m, $d_{12}=200$ m, $d_{21}=500$ m, path loss
exponent $\gamma=3.5$, and transmit power of $P_1=P_2=10$ dBm. The
per-stage sum quantization rate $L=8$ bits (i.e. the per-stage
quantization rate per antenna is $\frac{L}{K\times
N^2}=\frac{L}{8}=1$ bit). In (b), the ``*'', ``o'', etc represent
the simulation results of the proposed TICOQ and TVCOQ, while the
dashed line represents the analytical expression
$\mathcal{O}(2^{-\frac{L}{n}})$ (TICOQ) and $\bar
T\alpha^{\frac{\bar T-1}{2}}\mathcal{O}(2^{-\frac{L}{n}})$, with $\
\bar T\geq 1$ (TVCOQ) at fixed $L=8$ bits.}
    \label{Fig:vs_t}
\end{figure}

Fig. \ref{Fig:vs_t} (b) shows the tradeoff between convergence error
and message passing overhead (in terms of $\bar T$ at fixed $L$). As
the total number of iterations increases, the convergence error of
TVCOQ decreases, while the convergence error of the TICOQ and the
uniform quantizer fail to decrease. It is because that TICOQ has
steady state convergence error floor for any finite $L$ (i.e.
$\bar{E}^{\mathbf{w}}_{\text{block}}(\infty)
=\frac{1}{1-\alpha}\mathcal{O}(2^{-\frac{L}{n}})>0$ shown in Lemma
\ref{Lem:performace-tradeoff-invariant}), while the convergence
error of TVCOQ goes to 0 as $\bar T$ goes to infinity (i.e.
$\tilde{E}^{\mathbf{w}}_{\text{block}}(\bar T)=\bar
T\alpha^{\frac{\bar T-1}{2}}\mathcal{O}(2^{-\frac{L}{n}})\rightarrow
0$ as $\bar T\rightarrow \infty$ shown in Lemma
\ref{Lem:performace-tradeoff-variant}). Similarly, as the total
number of iterations increases, the sum throughput performance of
the TVCOQ improves but this is not the case for TICOQ and the
Baseline 2 as shown in Fig. \ref{Fig:vs_t} (a).

\section{Summary}\label{sec_summary}

In this paper, we study the convergence behavior of general
iterative function evaluation algorithms with quantized message
passing. We first obtain closed-form expressions of the  convergence
performance under quantized message passing among distributed nodes.
To minimize the effect of the quantization error on the convergence,
we propose the {\em time invariant convergence-optimal quantizer}
(TICOQ) and the {\em time varying convergence-optimal quantizer}
(TVCOQ). We found that the convergence error scales with the number
of bits for quantized message passing in the order of
$\frac{1}{1-\alpha}\mathcal{O}(2^{-\frac{L}{n}})$ and $\bar
T\alpha^{\frac{\bar T-1}{2}}\mathcal{O}(2^{-\frac{L}{n}})$ for TICOQ
and TVCOQ respectively. Finally, we illustrate using MIMO
interference game as example that the proposed designs achieve
significant gain in the convergence performance.

\begin{appendix}

\section*{Appendix A: Proof of Lemma \ref{Lem:contr-mapping-cov-error}}
First, we prove conclusion (a). By the update equation in
\eqref{eqn:contr-update-error}, the triangle inequality of norm and
the property of the contraction mapping, we have
$\|\mathbf{x}(t)-\mathbf{x}^*\|
=\|\mathbf{T}\big(\mathbf{x}(t-1)\big)+\mathbf{e}(t-1)-\mathbf{x}^*\|\leq\|\mathbf{T}\big(\mathbf{x}(t-1)\big)-\mathbf{x}^*\|+\|\mathbf{e}(t-1)\|
\leq
\alpha\|\mathbf{x}(t-1)-\mathbf{x}^*\|+\|\mathbf{e}(t-1)\|=\alpha\|\mathbf{T}\big(\mathbf{x}(t-2)\big)+\mathbf{e}(t-2)-\mathbf{x}^*\|+\|\mathbf{e}(t-1)\|
 \leq \cdots \leq
\alpha^t\|\mathbf{x}(0)-\mathbf{x}^*\|+E(t)$,
where $E(t)\triangleq\sum_{l=1}^t
\alpha^{l-1}\|\mathbf{e}(t-l)\|\stackrel{(1)}{=}\sum_{l'=0}^{l-1}\alpha^{t-l'-1}\|\mathbf{e}(l')\|\stackrel{(2)}{=}\alpha^{t-1}\sum_{l=0}^{t-1}\alpha^{-l}\|\mathbf{e}(l)\|$,
(1) is obtained by denoting $l'=t-l$, and (2) is obtained by
denoting $l=l'$. (b) is trivial. \textcolor{black}{Finally, we prove
conclusion (c). Since $\|\mathbf{e}(t)\|\leq \|\bar{\mathbf{e}}\|$
$\forall t$, we have $E(t)=\sum_{l=1}^t
\alpha^{l-1}\|\mathbf{e}(t-l)\|\leq \bar{E}(t)\triangleq\sum_{l=1}^t
\alpha^{l-1}\|\bar{\mathbf{e}}\|=\frac{1-\alpha^t}{1-\alpha}\|\bar{\mathbf{e}}\|$
and $\bar E(\infty)\triangleq \lim_{t\rightarrow \infty}\bar
E(t)=\frac{\|\bar{\mathbf{e}}\|}{1-\alpha}$. Given the limiting
error bound $\bar E(\infty)$, we know that $\exists T\ s.t. \
\forall t>T,\ \mathbf x(t) \in \mathbb S$. If $\mathbf
x=\boldsymbol{ \mathcal Q} \big( \mathbf T (\mathbf x)\big)$
$\forall \mathbf x \in \mathbb S$, then $\mathbf x(t) \to \mathbf
x(\infty)\in \mathbb S$. Thus, we obtain the sufficient condition
for convergence. On the other hand, if $\mathbf x \neq \boldsymbol{
\mathcal Q} \big( \mathbf T (\mathbf x)\big)$ $\forall \mathbf x \in
\mathbb S$, $\mathbf x(t)$ will not converge, but jumps among (at
least two) points in $\mathbb S$. Thus, we obtain the necessary
condition for convergence.}


\section*{Appendix B: Proof of Lemma \ref{Lem:Jacobi-Gauss-Seidel-cov-error}}
Jacobi scheme in \eqref{eqn:update-component-error} shares the
similar form as \eqref{eqn:contr-update-error}. Therefore, the proof
of Jacobi scheme is the same as that in Appendix A, except based on
weighted block-maximum norm.

Next, we prove the convergence for the Gauss-Seidel scheme under
quantized message passing. Let $\hat{\mathbf{x}}^k=\mathbf x$ for
$k=1$ and
$\hat{\mathbf{x}}^k=\big(\hat{\mathbf{S}}_1(\mathbf{x})+\mathbf{e}_1,\cdots,\hat{\mathbf{S}}_{k-1}
(\mathbf{x})+\mathbf{e}_{k-1},\mathbf{x}_k,\cdots,\mathbf{x}_K\big)$
for $2\leq k \leq K$. By the definition of weighted block-maximum
norm and the property of block-contraction $\mathbf{T}$, we have
\begin{align}
&\frac{\|\hat{\mathbf{S}}_k(\mathbf{x})-\mathbf{x}^*_k\|_k}{w_k}=\frac{\|\mathbf{T}_k(\hat{\mathbf{x}}^k)-\mathbf{T}_k(\mathbf{x}^*)\|_k}{w_k}\nonumber\\
&\leq
\|\mathbf{T}(\hat{\mathbf{x}}^k)-\mathbf{T}(\mathbf{x}^*)\|^{\mathbf{w}}_{\text{block}}
\leq  \alpha
\|\hat{\mathbf{x}}^k-\mathbf{x}^*\|^{\mathbf{w}}_{\text{block}}\nonumber\\
&\left\{\begin{array}{ll} \leq \alpha
\|\mathbf{x}-\mathbf{x}^*\|^{\mathbf{w}}_{\text{block}},&
k=1\\
=\alpha \max \Big\{
\max_{j<k}\frac{\|\hat{\mathbf{S}}_j(\mathbf{x})+\mathbf{e}_j-\mathbf{x}^*_j\|_j}{w_j},& \\
\quad \quad \quad \quad \ \max_{j\geq k}
\frac{\|\mathbf{x}_j-\mathbf{x}^*_j\|_j}{w_j} \Big\} , & 2\leq k
\leq K
\end{array} \right.\label{eqn:hat-s}
\end{align}
When $k=2$, by \eqref{eqn:hat-s}, we have
\begin{align}
&\frac{\|\hat{\mathbf{S}}_2(\mathbf{x})-\mathbf{x}^*_2\|_2}{w_2}\nonumber\\
\leq & \alpha \max \Big\{
\frac{\|\hat{\mathbf{S}}_1(\mathbf{x})+\mathbf{e}_1-\mathbf{x}^*_1\|_1}{w_1},\max_{j\geq
2} \frac{\|\mathbf{x}_j-\mathbf{x}^*_j\|_j}{w_j}
\Big\}\nonumber\\
\leq & \alpha \max \Big\{
\frac{\|\hat{\mathbf{S}}_1(\mathbf{x})-\mathbf{x}^*_1\|_1}{w_1}+\frac{\|\mathbf{e}_1\|_1}{w_1},
\, \max_{j\geq 2} \frac{\|\mathbf{x}_j-\mathbf{x}^*_j\|_j}{w_j}
\Big\} \nonumber\\
\leq & \alpha
\|\mathbf{x}-\mathbf{x}^*\|^{\mathbf{w}}_{\text{block}}+\alpha\|\mathbf{e}\|^{\mathbf{w}}_{\text{block}}\stackrel{\text{by
iteration}}{\Rightarrow}\nonumber
\end{align}
\begin{align}
&\frac{\|\hat{\mathbf{S}}_k(\mathbf{x})-\mathbf{x}^*_k\|_k}{w_k}\leq
\alpha
\|\mathbf{x}-\mathbf{x}^*\|^{\mathbf{w}}_{\text{block}}+\sum_{l=1}^{k-1}\alpha^l\|\mathbf{e}\|^{\mathbf{w}}_{\text{block}},
\forall k
\nonumber\\
\Rightarrow &
\|\hat{\mathbf{S}}(\mathbf{x})-\mathbf{x}^*\|^{\mathbf{w}}_{\text{block}}
\leq \alpha
\|\mathbf{x}-\mathbf{x}^*\|^{\mathbf{w}}_{\text{block}}\nonumber\\
&\quad \quad \quad \quad \quad \quad \quad \quad +
\frac{\alpha(1-\alpha^{K-1})}{1-\alpha}\|\mathbf{e}\|^{\mathbf{w}}_{\text{block}}\nonumber\\
\Rightarrow
&\|\mathbf{x}(t)-\mathbf{x}^*\|^{\mathbf{w}}_{\text{block}}\nonumber\\
=&\|\hat{\mathbf{S}}\big(\mathbf{x}(t-1)\big)+\mathbf{e}(t-1)-\mathbf{x}^*\|^{\mathbf{w}}_{\text{block}}\nonumber\\
\leq&
\alpha^t\|\mathbf{x}(0)-\mathbf{x}^*\|^{\mathbf{w}}_{\text{block}}+E^{\mathbf{w}}_{\text{block}}(t)
\nonumber
\end{align}
\textcolor{black}{where $E^{\mathbf{w}}_{\text{block}}(t)\triangleq
\frac{1-\alpha^K}{1-\alpha} \alpha^{t-1}\sum_{l=0}^{t-1}
\alpha^{-l}\|\mathbf{e}(l)\|^{\mathbf{w}}_{\text{block}}$. Since we
have shown
$\|\mathbf{x}(t)-\mathbf{x}^*\|^{\mathbf{w}}_{\text{block}}\leq
\alpha^t\|\mathbf{x}(0)-\mathbf{x}^*\|^{\mathbf{w}}_{\text{block}}+E^{\mathbf{w}}_{\text{block}}(t)$,
which is the same as the conclusion in Lemma
\ref{Lem:contr-mapping-cov-error} (a) except for the different norm
$\|\cdot\|^{\mathbf{w}}_{\text{block}}$ and the extra scalar
$\frac{1-\alpha^K}{1-\alpha}>1$ (indicating the additional error due
to the incremental nature of the Gauss-Seidel update) in
$E^{\mathbf{w}}_{\text{block}}(t)$, we can follow the similar steps
in Appendix A to obtain the conclusion for Gauss-Seidel scheme.}

\section*{Appendix C: Proof of Lemma \ref{Lem:SQ}}
First, we shall show that $\|\cdot\|^{\mathbf{w}}_{\text{block}}$ is
monotone (absolute).  Denote
$|\mathbf{x}|\triangleq(|x_1|,\cdots,|x_n|)$. We say that
$|\mathbf{x}|\leq |\mathbf{y}|$ if $|x_m|\leq |y_m|\, \forall m$.
Due to the monotonicity of $\|\cdot\|_k \ (\forall k)$, we have
$|\mathbf{x}|\leq |\mathbf{y}| \Leftrightarrow |\mathbf{x}_k|\leq
|\mathbf{y}_k|\, (\forall k) \Rightarrow \|\mathbf{x}_k\|_k\leq
\|\mathbf{y}_k\|_k\, (\forall k) \Rightarrow \max_k
\frac{\|\mathbf{x}_k\|_k}{w_k} \leq \max_k
\frac{\|\mathbf{y}_k\|_k}{w_k} \Leftrightarrow
\|\mathbf{x}\|^{\mathbf{w}}_{\text{block}}\leq
\|\mathbf{y}\|^{\mathbf{w}}_{\text{block}}$. Thus,
$\|\cdot\|^{\mathbf{w}}_{\text{block}}$ is monotone (absolute).
Next, we shall  show that each coordinate scalar TICOQ is uniform
quantizer. Given any $\mathbf{L}^s$ s.t.
\eqref{eqn:SQ-sum-rate-cons} is satisfied, by the monotonicity of
$\|\cdot\|^{\mathbf{w}}_{\text{block}}$, we can easily prove
$\min_{\boldsymbol{\mathcal
Q}^s(\mathbf{L}^s)}\|\bar{\mathbf{e}}\|^{\mathbf{w}}_{\text{block}}
\Leftrightarrow \min_{\mathcal Q_m^s(L_m^s)} \bar{e}_m (\forall m)$.
In other words, given any $L_m^s$, the coordinate scalar TICOQ
$\mathcal Q_m^{s*}(L_m^s)$ should minimize the worst-case error
$|\bar{e}_m|$ for the $m$-th coordinate of the input vector. Since
the uniform quantizer minimizes the worst-case error regardless of
the shape of the input pdf \cite{GrayQuantizationBook:1992}, each
coordinate scalar TICOQ $\mathcal Q_m^{s*}$ is a uniform quantizer.

\section*{Appendix D: Proof of Theorem \ref{Lem:sol-SQ-weighted} and Theorem \ref{Lem:sol-SQ-Lp}}

\textcolor{black}{When $\|\cdot\|_k$ is weighted maximum norm, the
objective function becomes
$\|\bar{\mathbf{e}}\|^{\mathbf{w}}_{\text{block}}=\max_k
\frac{\|\bar{\mathbf{e}}_k\|_k}{w_k}=\max_k \frac{\max_{m \in
\mathcal M_k}\frac{|\bar e_m|}{a_m}}{w_k}=\max_m\frac{|\bar
e_m|}{a_m(\sum_{k=1}^K w_k \mathbf I[m \in \mathcal M_k])}=\max_m
C_m 2^{-L^s_m}$. Therefore, we have $\min_{\boldsymbol{\mathcal
Q}^s}\|\bar{\mathbf{e}}\|^{\mathbf{w}}_{\text{block}}=\min_{\mathbf
L^s}\big(\max_m C_m 2^{-L_m^s}\big)$. By continuous relaxation and
equivalent transformation of minimax problems
\cite{Boydcvxbook:2004}, the minimax problem in Problem
\ref{Prob:TICOQ} is equivalent to the following problem (under
continuous relaxation), which is in epigraph form with optimization
variables $\{\bar L_m^s\},\tau$}:
\begin{align}
(\bar P^s): \min_{\{\bar L_m^s\},\tau} \ & \tau \nonumber\\
s.t. \ & C_m 2^{-\bar L_m^s} \leq \tau(1\leq m \leq n) \label{eqn:const-t}\\
 \ & \sum_{m=1}^n \bar L_m^s =L, \bar L_m^s\geq 0  (1\leq m \leq
n) \label{eqn:SQ-sum-rate-cons-relax}
\end{align}
$(\bar P^s)$ is a convex optimization problem. It can be easily
shown that the Slater's condition holds. Therefore, we shall get the
optimal solution to the relaxed problem through KKT conditions.
\textcolor{black}{The Lagrangian of ($\bar P^s$) is given by
$\mathcal L^s(\bar{\mathbf{L}}^s,\tau,\boldsymbol{\lambda}^s,
\boldsymbol{\nu}^s, \mu^s)= \tau+\sum_{m=1}^n \lambda^s_m (C_m
2^{-\bar L_m^s}-\tau) -\sum_{m=1}^n \nu^s_m \bar L_m^s + \mu^s
(\sum_{m=1}^n \bar L_m^s -L)$, where $\boldsymbol{\lambda}^s,
\boldsymbol{\nu}^s, \mu^s$ are the Lagrangian multipliers (LM).
$\bar{\mathbf{L}}^s$, $\tau$, $\boldsymbol{\lambda}^s,
\boldsymbol{\nu}^s, \mu^s$ are optimal  iff they satisfy the
following KKT conditions: (a) primal constraints:
\eqref{eqn:SQ-sum-rate-cons-relax},\eqref{eqn:const-t}; (b) dual
constraints: $\boldsymbol{\lambda}^s\succeq 0,
\boldsymbol{\nu}^s\succeq 0$; (c) complementary slackness:
$\lambda^s_m (C_m 2^{-\bar L_m^s}-\tau)=0 \ (\forall m),\ \nu^s_m
\bar L_m^s=0 \ (\forall m)$; (d) $ \frac{\partial \mathcal
L^s}{\partial \bar L_m^s}= -\ln(2) \lambda^s_m C_m 2^{-\bar
L_m^s}-\nu^s_m +\mu^s=0 \ (\forall m)$, $\frac{\partial \mathcal
L^s}{\partial \tau}=1-\sum_{m=1}^n \lambda^s_m=0$. Thus, $\forall
m$, we have
\begin{align}
\text{if }& \frac{\ln(2) \lambda^s_m C_m}{\mu^s}>1 :
\lambda^s_m=\frac{\mu^s}{\ln(2) \tau }, \nu^s_m=0,\nonumber\\
& \quad \quad  \bar L_m^{s*}=
\log_2(\frac{\ln(2) \lambda^s_m C_m}{\mu^s}),\tau=\frac{\mu^s}{\ln(2) \lambda^s_m }, \mu^s>0 \nonumber\\
\text{if }& \frac{\ln(2) \lambda^s_m C_m}{\mu^s}\leq 1 : \lambda^s_m
=0, \nu^s_m=\mu^s, \bar L_m^{s*}= 0,  \mu^s>0 \nonumber
\end{align}
where LMs $\{\lambda^s_m\}, \mu^s$ are chosen to satisfy
$\sum_{m=1}^n (\log_2\frac{\ln(2) \lambda^s_m C_m}{\mu^s})^{+}=L$
and $1-\sum_{m=1}^n \lambda^s_m=0$. Substitute
$\tau=\frac{\mu^s}{\ln(2) \lambda^s_m }$ into $\bar{L}_m^{s*}$, we
have $\bar{L}_m^{s*}=(\log_2\frac{C_m}{\tau})^{+}$, where $\tau$ is
chosen to satisfy $\sum_{m=1}^n (\log_2\frac{C_m}{\tau})^{+}=L$.
Furthermore, substituting  the relaxed solution  $\bar{L}_m^{s*}$
into the transformed problem $(\bar P^s)$, the optimal value of
$(\bar P^s)$ is given by $\tau$ and this is also the optimal value
$\|\bar{\mathbf{e}}^*\|^{\mathbf{w}}_{\text{block}}$ of the original
optimization Problem \ref{Prob:TICOQ} (under continuous relaxation)
due to the equivalence of the epigraph transformation. Next, we are
trying to prove that the rounding strategy in \eqref{eqn:sol-SQ} in
Theorem \ref{Lem:sol-SQ-weighted} is the optimal integer solution of
Problem \ref{Prob:TICOQ}. Suppose we round $\bar{L}_m^{s*}$ to
$\lfloor \bar{L}_m^{s*}\rfloor$ and let
$\delta^s_m=\bar{L}_m^{s*}-\lfloor \bar{L}_m^{s*}\rfloor$. Denote
$b_m=C_m 2^{-\lfloor \bar{L}_m^{s*}\rfloor}=C_m
2^{-(\bar{L}_m^{s*}-\delta^s_m)}$, i.e. $b_m=C_m$ if $C_m \leq \tau$
and $b_m=2^{\delta^s_m} \tau$ otherwise. The value of the objective
function in Problem \ref{Prob:TICOQ} is $\max_m b_m$.  Reducing
$\bar{L}_m^{s*}$ by $N+\delta^s_m$ ($ N \geq 1, \ N \in \mathbb
Z^+$) $\forall m$ for integer solution will lead to the value of
objective function greater than $\max_m b_m$. On the other hand,
since the optimal value of the original integer programming problem
is greater than the optimal value under continuous relaxation,
increasing $\bar{L}_m^{s*}$ by $N-\delta^s_m$ ($ N >1, \ N \in
\mathbb Z^+$) $\forall m$ will not help further reducing the value
of the objective function. Therefore, the optimal integer solution
$\{L_m^{s*}\}$ satisfies $\delta^s_m-1 \leq
\bar{L}_m^{s*}-L_m^{s*}\leq \delta^s_m $  and the rounding strategy
in \eqref{eqn:sol-SQ} is the optimal integer solution of Problem
\ref{Prob:TICOQ}. }

\textcolor{black}{When $\|\cdot\|_k$ is $L_p$ norm, the objective
function becomes
$\|\bar{\mathbf{e}}\|^{\mathbf{w}}_{\text{block}}=\max_k
\frac{\|\bar{\mathbf{e}}_k\|_k}{w_k}=\max_k \frac{(\sum_{m \in
\mathcal M_k}|\bar e_m|^p)^{\frac{1}{p}}}{w_k}=\max_k (\sum_{m \in
\mathcal M_k}\frac{|\bar e_m|^p}{\sum_{k=1}^K w_k^p \mathbf I[m \in
\mathcal M_k]})^{\frac{1}{p}}=\max_k (\sum_{m\in \mathcal M_k} C_m
2^{-pL_m^s})^{\frac{1}{p}}$. Therefore, we have
$\min_{\boldsymbol{\mathcal
Q}^s}\|\bar{\mathbf{e}}\|^{\mathbf{w}}_{\text{block}}=\min_{\mathbf
L^s}\big(\max_k (\sum_{m\in \mathcal M_k} C_m
2^{-pL_m^s})^{\frac{1}{p}}\big)$. Using similar continuous
relaxation and equivalent transformation of minimax problems
\cite{Boydcvxbook:2004}, the minimax problem in Problem
\ref{Prob:TICOQ} is equivalent (under continuous relaxation) to the
following problem, which is in epigraph form with optimization
variables $\{\bar L_m^s\},\tau$}:
\begin{align}
(\bar Q^s):  \min_{\bar L_m^s,\tau} \ & \tau \nonumber\\
s.t. \ & \sum_{m\in \mathcal M_k} C_m 2^{-p \bar L_m^s} \leq \tau \ (1\leq k \leq K) \label{eqn:const-t-Lp}\\
 & \text{constraint in}\eqref{eqn:SQ-sum-rate-cons-relax} \nonumber
\end{align}
$(\bar Q^s)$ is a convex optimization problem.
\textcolor{black}{Using similar argument as in the weighted maximum
norm case, the Lagrangian of ($\bar Q^s$) is given by $\mathcal
L^s(\bar{\mathbf{L}}^s,\tau,\boldsymbol{\lambda}^s,
\boldsymbol{\nu}^s, \mu^s)= \tau+\sum_{k=1}^K \lambda^s_k$ $
(\sum_{m\in \mathcal M_k} C_m 2^{-p \bar L_m^s} - \tau)
-\sum_{m=1}^n \nu^s_m \bar L_m^s + \mu^s (\sum_{m=1}^n \bar L_m^s
-L)$, where $\boldsymbol{\lambda}^s, \boldsymbol{\nu}^s, \mu^s$ are
the LMs. Using standard KKT conditions,
the optimal solution of $(\bar Q^s)$ is $\bar L_m^{s*}=
\frac{1}{p}\log_2 \Big(\frac{p\ln(2)(\sum_{k=1}^K \lambda^s_k
\mathbf{1}[m\in \mathcal M_k]) C_m}{\mu^s}\vee 1\Big)$ and
$\sum_{m\in \mathcal M_k} C_m 2^{-p \bar L_m^s} =\sum_{m\in \mathcal
M_k}\big(\frac{\mu^s}{p\ln(2) \lambda^s_k}\wedge C_m\big)=\tau$ if
$\lambda^s_k>0$, $x\vee a \triangleq \max\{x,a\}$, $x\wedge a
\triangleq \min\{x,a\}$, where the LMs $\{\lambda^s_k\}$, $\mu^s$
are chosen to satisfy $\sum_{m=1}^n \frac{1}{p}\log_2
\Big(\frac{p\ln(2)(\sum_{k=1}^K \lambda^s_k \mathbf{1}[m\in \mathcal
M_k]) C_m}{\mu^s}\vee 1\Big)=L$ and $\sum_{k=1}^K \lambda^s_k=1$.
Let $\tau_k=\frac{\mu^s}{p \ln(2) \lambda^s_k}$, then we have $\bar
L_m^{s*}= \frac{1}{p}\log_2 (\frac{ C_m}{\sum_{k=1}^K \tau_k \mathbf
I[m \in \mathcal M_k]}\vee 1)$, where 
$\{\tau_k\}$ and $\tau$ are constants related to the LMs
$\{\lambda^s_k\}$, $\mu^s$ of the constraints
\eqref{eqn:SQ-sum-rate-cons-relax},\eqref{eqn:const-t-Lp} in problem
$(\bar Q^s)$. They are chosen to satisfy $\sum_{k=1}^K \sum_{m\in
\mathcal M_k} \frac{1}{p}\log_2 (\frac{ C_m}{ \tau_k }\vee 1)=L$ and
$\frac{1}{\tau_k}\big(\sum_{m\in \mathcal M_k}(C_m \wedge
\tau_k)-\tau \big)=0$ ($\forall k$). Finally, substituting
$\bar{L}_m^{s*}$ into the transformed problem $(\bar Q^s)$, the
optimal value of $(\bar Q^s)$ is given by $\tau^{\frac{1}{p}}$ and
this is also the optimal value
$\|\bar{\mathbf{e}}^*\|^{\mathbf{w}}_{\text{block}}$ of the original
optimization Problem \ref{Prob:TICOQ} (under continuous relaxation)
due to the equivalence of the epigraph transformation.}

\section*{Appendix E: Proof of Lemma \ref{Lem:VQ} and \ref{Lem:sol-VQ-L2}}
Since $\|\mathbf{x}_k\|_k\leq \|\mathbf{y}_k\|_k\ (\forall k)
\Rightarrow\|\mathbf{x}\|^{\mathbf{w}}_{\text{block}}\leq
\|\mathbf{y}\|^{\mathbf{w}}_{\text{block}}$, given any $
\mathbf{L}^v$ s.t. \eqref{eqn:VQ-sum-rate-cons} is satisfied, we
have $\min_{\boldsymbol{\mathcal
Q}^v(\mathbf{L}^v)}\|\bar{\mathbf{e}}\|^{\mathbf{w}}_{\text{block}}\Leftrightarrow
\min_{\mathcal Q_k^v(L_k^v)} \|\bar{\mathbf{e}}_k\|_k \ (\forall
k)$.
The type of each component vector TICOQ $\mathcal Q_k^{v*}$ is
uniquely determined by the norm $\|\cdot\|_k$ on $\mathbb{R}^{n_k}$.
Furthermore, it is easy to prove that $\mathcal Q_k^v$ should be a
lattice quantizer to minimize the worst-case error
$\|\bar{\mathbf{e}}_k\|_k$. We shall discuss the two cases for $L_2$
norm  and weighted maximum norm separately.

\begin{itemize}
\item $\|\cdot\|_k$ ($1\leq k \leq K$) is  $L_2$ norm: 
The covering problem asks for the thinnest covering of
$\mathbb{R}^{n_k}$ dimensional space with overlapping spheres, i.e.
minimizes covering radius (circumradius of the Voronoi cell)
$\rho_k=\|\bar{\mathbf{e}}_k\|_k=\big(\sum_{m \in \mathcal M_k}
|\bar{e}_m|^2\big)^{\frac{1}{2}}$\cite{ConvwaySloaneLatticebook:1999}.
Therefore, each component vector TICOQ $\mathcal Q_k^{v*}$
minimizing the worst-case error is the thinnest lattice for the
covering problem.
\item $\|\cdot\|_k$ ($1\leq k \leq K$) is  weighted maximum norm:  Given any $ L_k^v$,
we have $\min_{\mathcal Q_k^v(L_k^v)}
\|\bar{\mathbf{e}}_k\|^{\mathbf{a}_k}_{\infty} =\min_{\mathcal
Q_k^v(L_k^v)} \max_{m \in \mathcal M_k} \frac{|\bar{e}_m|}{a_m}$. It
can be easily shown that each face of the Voronoi cell for the
weighted maximum norm is $(n_k-1)$- dimensional hyperplane parallel
to a coordinate axis in the $n_k$ dimensional space. Therefore, it
is equivalent to the scalar quantization of each coordinate $x_m$ of
the input block component $\mathbf x_k$ with different scalar
quantizers, i.e. $\mathcal Q_k^{v*}=(\mathcal Q_m^{s*})_{m\in
\mathcal M_k}$.
\end{itemize}

Next, we shall show the optimal solution for the dual lattice
quantizer. For $A^*_{n_k}$, the covering radius is
$R_{n_k}=\sqrt{\frac{n_k(n_k+2)}{12(n_k+1)}}$, the volume of the
fundamental region is $\sqrt{\frac{1}{n_k+1}}$. The volume of
bounded region $\boldsymbol{\mathcal{X}}_k$ is $V^B_k=\prod_{m \in
\mathcal M_k}|\mathcal X_m|$. Therefore, the worst case error of the
vector TICOQ with quantization rate $L_k^v$ is given by
$\|\bar{\mathbf{e}}_k\|_k=\Big(\frac{V^B_k}{2^{L^v_k}\sqrt{\frac{1}{n_k+1}}}\Big)^{\frac{1}{n_k}}R_{n_k}=\Big(\frac{\prod_{m
\in \mathcal M_k}|\mathcal
X_m|}{\sqrt{\frac{1}{n_k+1}}}\Big)^{\frac{1}{n_k}}\sqrt{\frac{n_k(n_k+2)}{12(n_k+1)}}2^{-\frac{L^v_k}{n_k}}$\footnote{Note
that boundary effect is ignored here. The performance loss is
negligible when $L$ is large, which is easily satisfied in most of
the cases we are interested in.}.
The covering radius measured by $L_p$  norm ($p>2$) can be proved to
be less than $R_{n_k}=\sqrt{\frac{n_k(n_k+2)}{12(n_k+1)}}$, which is
the covering radius measured by $L_2$ norm
. Thus, the worst case error is also less than
$\|\bar{\mathbf{e}}_k\|_k$ given above. Therefore, in general, we
can apply $A^*_n$ quantizer for VQ case when $\|\cdot\|_k$ is $L_p$
norm ($p\geq 2$) and consider TICOQ design for VQ cased based on
$A^*_n$ quantizer.

Problem \ref{Prob:TICOQ} for $A^*_n$ quantizer is equivalent to
$\min_{\mathbf L^v} \big(\max_k D_k 2^{-\frac{L^v_k}{n_k}}\big) s.t.
\eqref{eqn:VQ-sum-rate-cons}$. Similar to Appendix D, the optimal
solution under continuous relaxation is $\bar{L}_k^{v*}=n_k(\log_2
\frac{D_k}{\tau})^{+}$, \textcolor{black}{where $\tau$ is a constant
related to the LM and is chosen to satisfy} $\sum_{m=1}^n n_k(\log_2
\frac{D_k}{\tau})^{+}=L$. For $n_1=\cdots=n_K$, \textcolor{black}{we
can use similar argument as in Appendix D to show that the rounding
method in \eqref{eqn:sol-VQ} is optimal  integer solution to Problem
\ref{Prob:TICOQ} (VQ case).}



\section*{Appendix F: Proof of Lemma \ref{Lem:performace-tradeoff-invariant}}
\textcolor{black}{We try to find $L'$ in the following three cases
s.t. when $L\geq L'$, we have $C_m\geq\tau$ ($\forall m$), $C_m \geq
\sum_{k=1}^K \tau_k \mathbf I[m \in \mathcal M_k]$ ($\forall m$) and
$D_k \geq \tau$ ($\forall k$), separately (to obtain the closed-form
optimal value
$\|\bar{\mathbf{e}}^*\|^{\mathbf{w}}_{\text{block}}$).}
Specifically, we have
\begin{itemize}
\item SQ ($\|\cdot\|_k$ is weighted maximum norm) in Theorem \ref{Lem:sol-SQ-weighted} (same as VQ ($\|\cdot\|_k$ is weighted maximum norm)):
\textcolor{black}{$C_m\geq\tau \ (\forall m)\Leftrightarrow
\bar{L}_m^s=(\log_2 \frac{C_m}{\tau})^{+}=\log_2 \frac{C_m}{\tau}$
($\forall m$). Since $\sum_{m=1}^n \log_2 \frac{C_m}{\tau}
=L\Rightarrow
\|\bar{\mathbf{e}}^*\|^{\mathbf{w}}_{\text{block}}=\tau=2^{\frac{1}{n}\sum_{m=1}^n
\log_2 C_m-\frac{L}{n}}=\mathcal{O}(2^{-\frac{L}{n}})$, we have
$C_m\geq\tau \ (\forall m) \Leftrightarrow \min_m C_m \geq
2^{\frac{1}{n}\sum_{m=1}^n \log_2 C_m-\frac{L}{n}} \Leftrightarrow
L\geq  \sum_{m=1}^n \log_2 C_m-n \log_2 (\min_m C_m)\triangleq L'$.}

\item SQ ($\|\cdot\|_k$ is $L_p$ norm) in Theorem \ref{Lem:sol-SQ-Lp}:
\textcolor{black}{$C_m \geq \sum_{k=1}^K \tau_k \mathbf I[m \in
\mathcal M_k] \ (\forall m) \Leftrightarrow \bar L_m^s=
\frac{1}{p}\log_2 (\frac{ C_m}{\sum_{k=1}^K \tau_k \mathbf I[m \in
\mathcal M_k]}\vee 1)=\frac{1}{p}\log_2 (\frac{ C_m}{\sum_{k=1}^K
\tau_k \mathbf I[m \in \mathcal M_k]})$ ($\forall m$). Since
$\sum_{m\in \mathcal M_k} C_m 2^{-p \bar L_m^s} =n_k
\tau_k=\tau\Rightarrow \sum_{m=1}^n
\bar{L}_m^s=\frac{1}{p}\sum_{m=1}^n \log_2 \frac{\tilde C_m}{\tau}=L
\Rightarrow
\|\bar{\mathbf{e}}^*\|^{\mathbf{w}}_{\text{block}}=\tau^{\frac{1}{p}}=2^{\frac{1}{p}\big(\frac{1}{n}\sum_{m=1}^n
\log_2 (\tilde
C_m)-\frac{pL}{n}\big)}=\mathcal{O}(2^{-\frac{L}{n}})$. Similarly,
we have $L'=\sum_{m=1}^n \log_2\big( \tilde C_m \big)-n \log_2 \big(
\min_m \tilde C_m\big) $.}

\item VQ ($A^*_n$ quantizer) in Theorem \ref{Lem:sol-VQ-L2}:
\textcolor{black}{$D_k \geq \tau \ (\forall k) \Leftrightarrow
\bar{L}_k^v=n_k(\log_2 \frac{D_k}{\tau})^{+}=n_k\log_2
\frac{D_k}{\tau}$ ($\forall k$). Since $\sum_{k=1}^K n_k\log_2
\frac{D_k}{\tau} =L\Rightarrow
\|\bar{\mathbf{e}}^*\|^{\mathbf{w}}_{\text{block}}=\tau=2^{\frac{1}{n}\sum_{k=1}^K
n_k \log_2 D_k-\frac{L}{n}}=\mathcal{O}(2^{-\frac{L}{n}})$.
Similarly, we have $L'= \sum_{k=1}^K \log_2 D_k-K \log_2 (\min_k
D_k) $.}
\end{itemize}

\section*{Appendix G: Proof of Theorem \ref{Lem:sol-TVCOQ-masterp} and Lemma \ref{Lem:performace-tradeoff-variant}}
\textcolor{black}{First, we shall find the requirement for $L$ s.t.
each subproblem (under continuous relaxation) has closed-form
$\|\tilde{\mathbf{e}}_t^*\big(L(t)\big)\|^{\mathbf{w}}_{\text{block}}$
(to obtain the closed-form objective function of the TVCOQ master
problem). By Appendix F, to obtain closed-form
$\|\tilde{\mathbf{e}}_t^*\big(L(t)\big)\|^{\mathbf{w}}_{\text{block}}$,
we require $L(t)\geq L'$ ($ 0 \leq t \leq \bar T-1$). Under this
assumption, we have
$\|\tilde{\mathbf{e}}_t^*\big(L(t)\big)\|^{\mathbf{w}}_{\text{block}}=\eta
2^{-\frac{L(t)}{n}}$, where
\begin{align}
\eta=\left\{
\begin{array}{ll}
2^{\frac{1}{n}\sum_{m=1}^n \log_2 C_m}, &
\text{WM norm (SQ)} \\
2^{\frac{1}{pn}\sum_{m=1}^n \log_2 (\tilde
C_m)}, & \text{$L_p$ norm (SQ)}\\
2^{\frac{1}{n}\sum_{k=1}^K n_k \log_2 D_k}, & \text{dual lattice
(VQ)}
\end{array} \right..
\end{align}
Therefore, the objective function in Problem
\ref{Prob:TVCOQ-masterp} becomes $\eta \alpha^{\bar
T-1}\sum_{t=0}^{\bar T-1}\alpha^{-t}2^{-\frac{L(t)}{n}}$ and Problem
\ref{Prob:TVCOQ-masterp} is equivalent to $\min_{\{L(t)\}}
\sum_{t=0}^{\bar T-1} \alpha^{-t} 2^{-\frac{L(t)}{n}} s.t.
\eqref{eqn:TV-per-stage-sum-rate-allo}$. By continuous relaxation
and standard convex optimization techniques (similar to Appendix D),
we have the optimal solution (under continuous relaxation) $\bar
L^*(t)=n \log_2(\frac{\alpha^{-t}\ln 2 }{n \mu})^+$. Since $L(t)\geq
L' \ (\forall t)$, $\bar L^*(t)=n \log_2(\frac{\alpha^{-t}\ln 2 }{n
\mu})$. $\sum_{t=0}^{\bar T-1} \bar L(t)=n \sum_{t=0}^{\bar
T-1}\big( \log_2(\ln 2)-t \log_2 (\alpha)-\log_2(n \mu)\big)=nt
\log_2(\ln 2) -n \frac{(\bar T-1)\bar T}{2} \log_2 (\alpha)
-nt\log_2(n \mu) =\bar T L \Rightarrow \bar L(t)=n \frac{\bar
T-1}{2} \log_2 (\alpha)- nl \log_2(\alpha)+L$. Since $L(t)$
increases with $t$, to satisfy $L(t)\geq L' \ (\forall t)$, we
require $L(0)=n \frac{\bar T-1}{2} \log_2 (\alpha)+L \geq
L'\Rightarrow L\geq L'-n \frac{\bar T-1}{2} \log_2 \alpha $.
Therefore, when $L\geq L'-n \frac{\bar T-1}{2} \log_2 \alpha $, we
have $\bar L^*(t)=n \log_2(\frac{\alpha^{-t}\ln 2 }{n \mu})$ and
$\alpha^{-t}\|\tilde{\mathbf{e}}_t^*(\bar
L^*(t))\|^{\mathbf{w}}_{\text{block}}=\eta\cdot
\alpha^{-t}\frac{n\mu}{\alpha^{-t}\ln 2}=\eta \cdot \frac{n\mu}{\ln
2}$. Similar to Appendix D, the rounding policy \textcolor{black}{in
\eqref{eqn:sol-masterp-int}} can be \textcolor{black}{shown} to be
optimal.}


\textcolor{black}{Next, we shall analyze the tradeoff between
convergence error and message passing overhead for $L\geq L'-n
\frac{\bar T-1}{2} \log_2 \alpha$ ($L \in \mathbb Z^+$). Since it
has been shown that $\bar L(t)=n \frac{\bar T-1}{2} \log_2 (\alpha)-
nl \log_2(\alpha)+L$, we have $\eta\alpha^{\bar T-1}\sum_{t=0}^{\bar
T-1}\alpha^{-t}2^{-\frac{L(t)}{n}}=\eta\alpha^{\bar
T-1}\sum_{t=0}^{\bar T-1}\alpha^{-t}\big( \alpha^{-\frac{\bar
T-1}{2}}\cdot \alpha^t \cdot 2^{-\frac{L}{n}}\big)=\bar T
\alpha^{\frac{\bar T-1}{2}}\mathcal
O(2^{-\frac{L}{n}})\Rightarrow\tilde E^{\mathbf{w}}_{\text{block}}
(\bar T)=\bar T \alpha^{\frac{\bar T-1}{2}}\mathcal
O(2^{-\frac{L}{n}})$.}

\end{appendix}

\bibliographystyle{IEEEtran}
\bibliography{IEEEabrv,cuiying-bib}

\begin{biography}[{\includegraphics[width=1.0in,height=1.1in,clip,keepaspectratio]{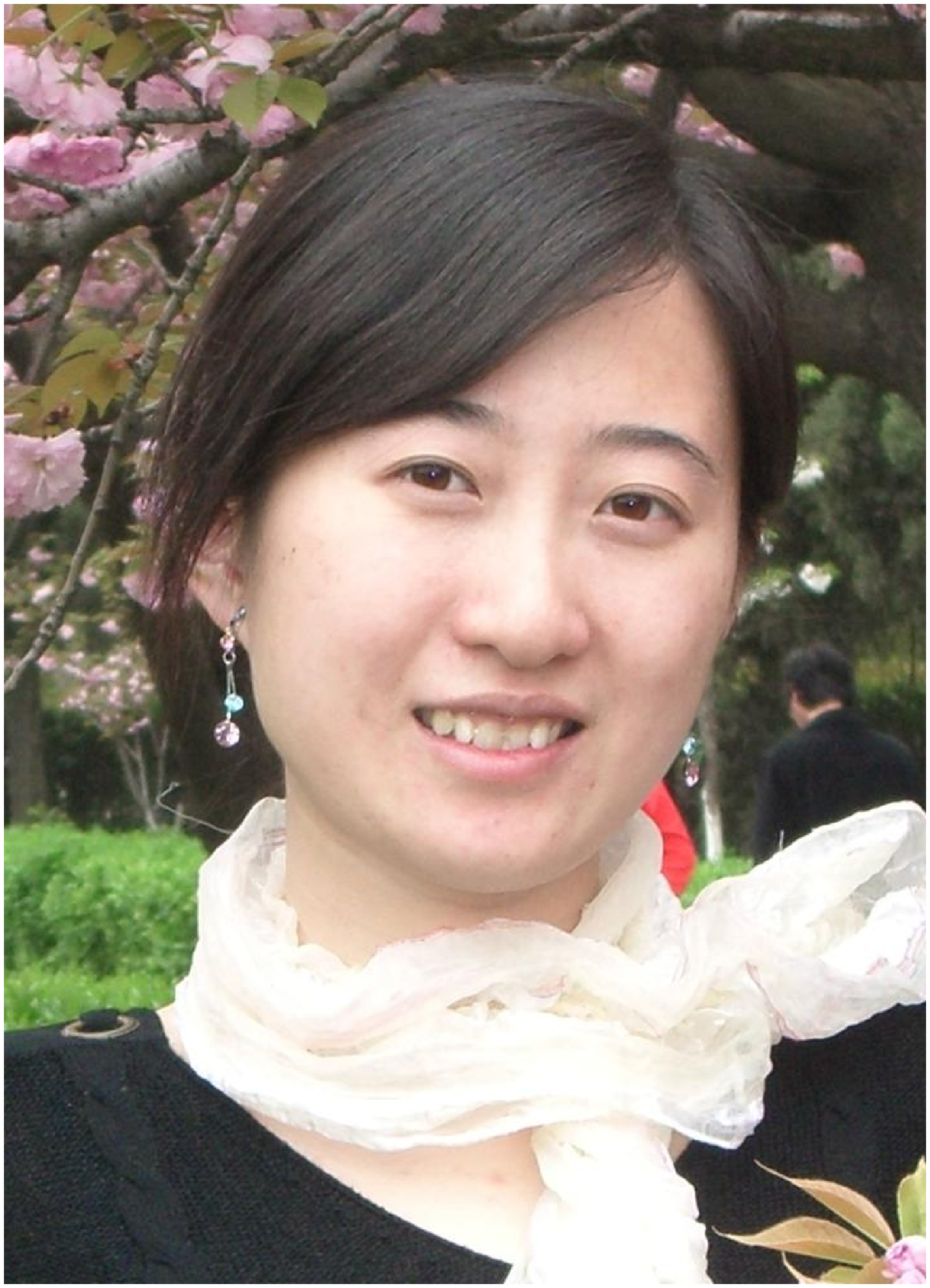}}]{Ying~Cui}
received B.Eng degree (first class honor) in Electronic and
Information Engineering, Xi¡¯an Jiaotong University, China in 2007.
She is currently a Ph.D candidate in the Department of ECE, the Hong
Kong University of Science and Technology (HKUST). Her current
research interests include cooperative and cognitive communications,
delay-sensitive cross-layer scheduling as well as stochastic
approximation and Markov Decision Process.
\end{biography}
\begin{biography}[{\includegraphics[width=1.0in,height=1.1in,clip,keepaspectratio]{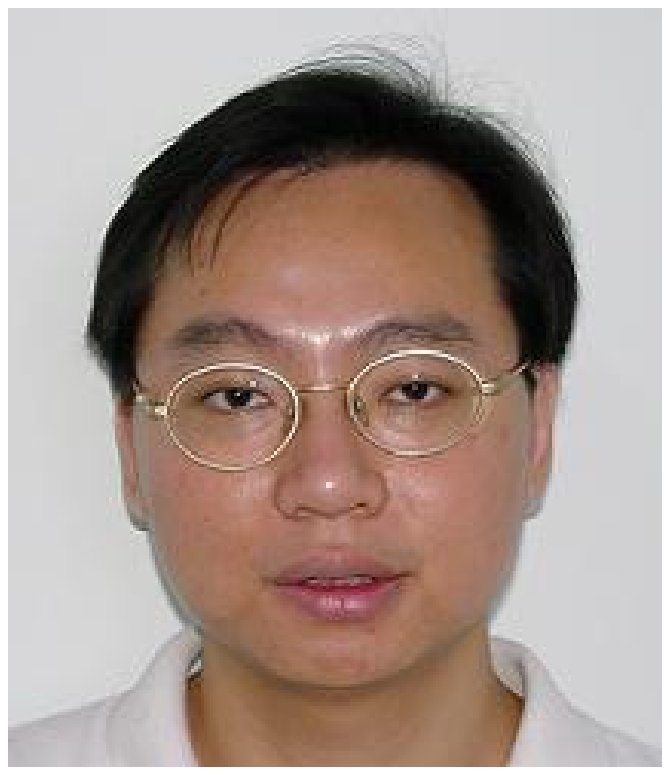}}]{Vincent~K.~N.~Lau}
obtained B.Eng (Distinction 1st Hons) from the University of Hong
Kong in 1992 and Ph.D. from Cambridge University in 1997. He was
with PCCW as system engineer from 1992-1995 and Bell Labs - Lucent
Technologies as member of technical staff from 1997-2003. He then
joined the Department of ECE, HKUST as Associate Professor. His
current research interests include the robust and delay-sensitive
cross-layer scheduling, cooperative and cognitive communications as
well as stochastic approximation and Markov Decision Process.
\end{biography}

\end{document}